\documentclass
[twocolumn,a4paper,showkeys,superscriptaddress]
{revtex4}
\usepackage{amsfonts}
\usepackage{amssymb}
\usepackage{amsmath}
\usepackage{graphicx}
\usepackage{natbib}
\usepackage{times}
\usepackage{epsfig}
\usepackage{float}
\usepackage{flafter}
\usepackage{psboxit}
\usepackage{fancyheadings}

\bibstyle{abbrvnat}
 
\begin{document}
\bibliographystyle{abbrvnat}

\title{
    Life cycle of a minimal protocell --- 
    a dissipative particle dynamics (DPD) study
}
\author{Harold Fellermann}
\affiliation{\textit{ICREA-Complex  Systems   Lab,  Universitat
Pompeu  Fabra  (GRIB),  Dr   Aiguader  80,  08003  Barcelona,  Spain}}
\thanks{Authors for correspondence: harold.fellermann@upf.edu; 
\\+34 935422834; Fax: +34 932213237 or \\steen@lanl.gov; +1-505-665-0052}

\author{Steen Rasmussen}   
\affiliation{\textit{Self-Organizing Systems, EES-6 MS-D462, Los Alamos National Laboratory, 
Los Alamos NM 87545, USA }}
\affiliation{\textit{Santa Fe  Institute, 1399  Hyde  Park Road,
Santa Fe NM 87501, USA}}

\author{Hans-Joachim Ziock}   
\affiliation{\textit{Self-Organizing Systems, EES-6 MS-D462, Los Alamos National Laboratory, 
Los Alamos NM 87545, USA }}

\author{Ricard V. Sol\'e}   
\affiliation{\textit{ICREA-Complex  Systems   Lab,  Universitat
Pompeu Fabra (GRIB), Dr Aiguader 80, 08003 Barcelona, Spain}}
\affiliation{\textit{Santa Fe  Institute, 1399  Hyde  Park Road,
Santa Fe NM 87501, USA}}

%
%
%
%
%
%
%
%
%
\begin{abstract}
Cross-reactions and other systematic issues generated by the coupling
of functional chemical subsystems pose the largest challenge for 
assembling a viable protocell in the laboratory. Our current work 
seeks to identify and clarify such key issues as we represent and 
analyze in simulation a full implementation of a minimal protocell.
Using a 3D dissipative particle dynamics (DPD) simulation method we 
are able to address the coupled diffusion, self-assembly, and 
chemical reaction processes, required to model a full life cycle of 
the protocell, the protocell being composed of coupled genetic, 
metabolic, and container subsystems. Utilizing this minimal 
structural and functional representation of the constituent 
molecules, their interactions, and their reactions, we identify and 
explore the nature of the many linked processes for the full 
protocellular system. Obviously the simplicity of this simulation 
method combined with the inherent system complexity prevents us 
from expecting quantitative simulation predictions from these 
investigations. However, we report important findings on systemic 
processes, some previously predicted, and some newly discovered, as 
we couple the protocellular self-assembly processes and chemical 
reactions. For example, our simulations indicate that the container 
stability is significantly affected by the amount of oily precursor 
lipids and sensitizers and affect the partition of molecules in the 
container division process. Also a continuous supply of oily lipid 
precursors to the protocell environment at a very slow rate will 
pulse the protocellular loading (feeding) as oil blobs will form in 
water and whole blobs will be absorbed at one time. By orchestrating 
the precursor injection rate compared to diffusion, precursor 
self-assembly, protocell concentration, etc., an optimal size 
resource package can be spontaneously generated. Replication of the 
amphiphilic genes is better on the surface of a micelle with a 
substantial oil core (loaded micelle) than on a regular micelle due 
to the higher aggregate stability. Also replication of amphiphilic 
genes (genes with lipophilic backbones) in bulk water can be 
inhibited due to their tendency to form aggregates. Further the 
template directed gene ligation rate depends not only on the 
component monomers but also on the sequence of these monomers in 
the template. 
\end{abstract}

\keywords{
artificial life, minimal protocell, diffusion, 
self-assembly, chemical reactions, dissipative particle dynamics
}
\maketitle

\section{Introduction}

The twilight zone that separates nonliving matter from life involves 
the assembly of and cooperation among different sub-components, which 
we can identify as metabolism, information, and compartment. None of 
these ingredients are living and none of them can be ignored when 
looking at life as a whole. When assembled appropriately in a 
functional manner, their systemic properties constitute minimal life.

Understanding the tempo and mode of the transition from nonliving to 
living matter requires a considerable effort of simplification 
compared to modern life. Cells as we know them in our current 
biosphere are highly complex. Even the simplest, parasitic cellular 
forms involve hundreds of genes, complex molecular machineries of 
energy exchange and intricate membrane structures \citep{Alb:2002}. 
Such modern organisms are presumably far away from the initial simple 
forms of cellular life that inhabited our planet a long time ago, 
whose primitive early cousins we are now attempting to assemble in the 
laboratory.

Several complementary designs of protocells have been proposed that 
differ in the actual coupling between their various internal 
components \citep{Lui:1994,Poh:2002,Gan:2003,Ras:2003,Han:2004}. One 
particularly important problem here, beyond the specific physical and 
chemical difficulties associated with the assembly of these 
protocells, is the problem of modeling the coupling of the possible 
kinetic and structural scenarios that lead to a full cell cycle. None 
of the current proposed designs has yet been formulated in a full 
mathematical model that in a 3D simulation is able to generate the 
possible outcomes of a successful coupling between the three prime 
components: the genes, the metabolism, and the container. We believe 
that a physically well-grounded modeling approach can provide critical
insight into what can be expected from a coupled set of structures 
and reactions, how the nano-scale stochasticity can jeopardize 
appropriate molecular interactions or even what are the effects of 
molecular information carriers in helping accurate replication to 
occur. In this paper we present such a minimal 3D model that in 
connection with ongoing experimental efforts is aimed at assembling 
and understanding a new class of nanoscale-sized protocells: the so 
called {\em Los Alamos Bug}.

\begin{figure}[bt]
  \centering
  \includegraphics[width=\columnwidth]{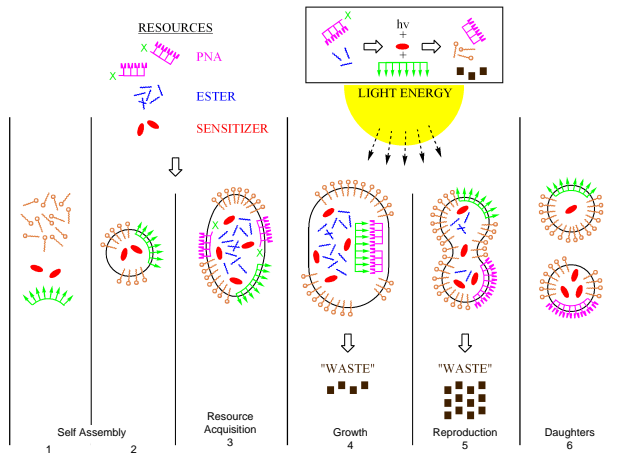}
  \caption{
    Schematic of the life cycle of the Los Alamos Bug: The system 
    consists of surfactants, sensitizers, and a biopolymer that acts 
    as a genome (1). The surfactants spontaneously self-assemble into 
    a micellar container within which the sensitizer resides while 
    the biopolymer sticks at the surface of the container---this 
    forms a complete protocell (2). Resources (genomic oligomers, 
    sensitizers and surfactant precursors in the form of esters) are 
    added to the system and get incorporated into the container (3). 
    The existing information carrier acts as a template for supplied 
    oligomers to hybridize and effectively replicate the genome. 
    Light energy is used to convert the surfactant precursor and the 
    oligomer precursors into actual surfactant, oligomers and waste. 
    The container grows as new surfac­tants are produced (5). Once the 
    container reaches a critical size, it becomes unstable and 
    divides into two daughter cells. This completes the life cycle of 
    the protocell (6).
  }
  \label{fig_cartoon}
\end{figure}

In the Los Alamos bug, the container is built of amphiphilic 
surfactants. Due to a their interaction with water, the surfactants 
spontaneously self-assemble into micelles with the hydrophobic end of
the surfactant molecules in the interior of the micelles and their 
hydrophilic ends in contact with the surrounding water.  The 
interactions between the micelle and the other components of the Los 
Alamos bug, namely the photosensitizer, the genome, and the container 
precursors, allow the micelles to host these other components. 

The genomic biopolymer (possibly decorated with hydrophobic anchors) 
is also an amphiphile and due to the specific nature of its 
interactions with water and the micelle, it will tend to reside at the 
surface of the micelle (see figure \ref{fig_cartoon}.2). The 
sensitizer is a hydrophobic molecule and will therefore reside in the
interior of the micelle. Once self-assembled, the protocell aggregate 
is ``fed'' with precursor molecules for the surfactants (oily esters), 
sensitizers and genomic precursor oligomers. As surfactant precursors 
are hydrophobic they will 
agglomerate inside the proto-organism and form a hydrophobic core 
(figure \ref{fig_cartoon}.3). Light energy is used by the metabolism 
to transform precursors into new building blocks (surfactants and 
oligomers) of the protocell. The genomic oligomers that are 
complementary with particular stretches of the template strand will 
hybridize with it (figure \ref{fig_cartoon}.4). The fully hybridized 
template/oligomers complex, which now only has hydrophobic elements 
exposed, will move into the interior of the container where 
polymerization of the oligomers occurs followed at some later time by 
a random dissociation of the fully polymerized double-stranded genome 
into two single-stranded templates that move back to the surface. 
This process could also be enhanced by a gentle temperature cycle near
the gene duplex melting point.

As surfactant precursors are digested, the core volume of the 
protocell decreases while, at the same time, new surfactants are 
produced. The resulting change in the surface to volume ratio causes
the micelle to become unstable (figure \ref{fig_cartoon}.5), until it 
finally splits into two daughter 
cells (figure 1.6). Assuming that components of the growing parent
micelle are appropriately distributed upon division,
the two daughter cells will be replicates of the original organism, 
thus completing the protocell cycle.

\begin{figure}[hbt]
  \centering
  \includegraphics[width=.6\columnwidth]{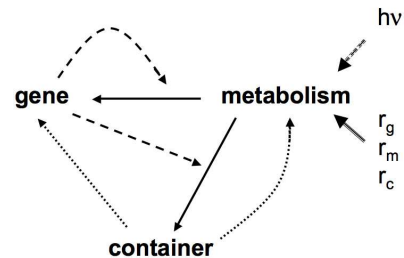}
  \caption{
    Functional coupling between container, metabolism and genome. 
    Note how the gene catalyses (dashed arrows) the metabolic 
    production (solid arrows) of both gene and container building 
    blocks. The container ensures a high local concentrations 
    (proximity) and facilitates thermodynamic reaction conditions 
    (dotted arrows) of both the metabolic molecules and the 
    amphiphilic replicator polymers. The free energy is provided by 
    light ($h\nu$) and the provided resources are precursor lipids 
    $r_c$, precursor gene oligomeres $r_g$, as well as sensitizers 
    $r_m$. 
  }
  \label{fig_coupling}
\end{figure}

In the above setup, the container, genome and metabolism are coupled 
in various ways. Obviously, both the replication of the container and 
replication of the genome depend on a functioning metabolism, as the 
latter provides building blocks for aggregate growth and reproduction. 
In addition to that, the container also has a catalytic influence on
the replication of both the metabolic elements and the genome: the 
micellar structure provides a compartment which brings precursors, 
sensitizers and nucleic acids in close vicinity, thereby increasing 
local concentrations and thus metabolic turnover. Furthermore, the 
micellar interface catalyzes the hybridization of the informational
polymer with its complementary oligomer. Once the hybridized complex 
enters the ``water-poor/free'' interior of a micelle, the 
thermodynamics should change sufficiently to allow a dehydration 
reaction to occur whereby the oligomers become polymerized. 
Alternatively the water-lipid interface could either itself act as a 
ligation catalyst or the addition of simple amphiphilic catalysts 
could facilitate the gene polymerization process. Last, but 
not least, the nucleic acid catalyzes the metabolism, which otherwise
is extremely slow. A summary of the subsystem coupling is shown in 
Fig. \ref{fig_coupling}. 

\section{The model}
Dissipative particle dynamics (DPD) is a mesoscale simulation method 
introduced by \citeauthor*{Hoo:1992} in \citeyear{Hoo:1992}. The 
method has been improved as a result of various theoretical support,
revision, and expanded capabilities 
\citep{Esp:1995,Mar:1998,Gro:1997}, and has been applied to a number 
of biological systems such as membranes \citep{Ven:1999,Gro:2001}, 
vesicles \citep{Yam:2002,Yam:2003}, and micelles 
\citep{Gro:2000,Yua:2002}. Also chemical reactions have been 
incorporated into the DPD method \citep{Bed:2005,Buc:2006}.  
In the context of protocells, DPD has recently been applied to study a 
self-replicating micellar system \citep{Fel:2006}. The DPD formalism 
used in this work is the revised version from Groot and Warren 
\citep{Gro:1997} that has become the de facto standard of DPD.

\subsection{Dissipative particle dynamics}
\label{sec_dpd}
A DPD simulation consists of a set of $N$ particles located in 
three-dimensional continuous space with Euclidean metrics. These 
particles are not individual atoms but represent several water 
molecules or beads in a polymer chain.
Each particle $i$ has a position ${\mathbf r}_i$, mass $m_i$ and 
momentum ${\mathbf q}_i$, from which one can derive its velocity 
${\mathbf v}_i = {\mathbf q}_i/m_i$. Its motion is determined by a 
force field ${\mathbf F}_i$ through Newton's second law of motion:
\begin{equation}
    \frac{d^2{\mathbf r}_i}{dt^2}(t) 
        = \frac{1}{m_i} {\mathbf F}_i\left(\mathbf r_i(t)\right)
    \label{eqn_dynamics}
\end{equation}

The force acting on particle $i$ can be decomposed into pair-wise 
interactions, which respectively are the sum of three different 
components---a conservative, a dissipative and a random one:
\begin{equation}
    \mathbf F_i = \sum_{j\not=i} \mathbf F_{ij}
        = \sum_{j\not=i} \left( \mathbf F_{ij}^C + \mathbf F_{ij}^D
        + \mathbf F_{ij}^R \right) \mathbf{ ,}
    \label{eqn_forces}
\end{equation}
where $\mathbf{F}^C$, $\mathbf{F}^D$ and $\mathbf{F}^R$ are defined by
\begin{eqnarray}
    \mathbf F_{ij}^C & = & -\boldsymbol \nabla \phi_{ij} \\
    \mathbf F_{ij}^D & = & -\eta \omega^D(r_{ij})
        \left(\mathbf n_{ij}\cdot
        \mathbf v_{ij}\right)\mathbf n_{ij} \label{eqn_dissipative} \\
    \mathbf F_{ij}^R & = & \sigma \omega^R(r_{ij})
        \xi_{ij}\mathbf n_{ij} \label{eqn_random}
\end{eqnarray}
For each particle pair $(i,j)\:$ 
$\mathbf r_{ij} = \mathbf r_i - \mathbf r_j$ is the relative position, 
$r_{ij} = \left|\mathbf r_{ij}\right|$ the center-to-center distance, 
and 
$\mathbf v_{ij} = \mathbf v_i - \mathbf v_j$ the relative velocity. We 
denote with $\mathbf n_{ij} = \mathbf r_{ij}/r_{ij}$ the (unit)
direction between the two particles. A detailed discussion of the
different forces $\mathbf F_{ij}^X$ now follows:

The conservative force ${\mathbf F}_{ij}^C$ is expressed in the usual 
way as the negative gradient of a potential 
$\phi_{ij} = V_{ij} = V(r_{ij})$. In most DPD simulations, a pure 
repulsive soft core potential of the form
\begin{equation}
    V_{ij}(r) = \left\{
    \begin{array}{cl}
        \frac{a_{ij}}{2}(r-r_c)^2 & \mbox{ if } r<r_c \\
        0 & \mbox{ otherwise}
    \end{array}
    \right.
    \label{eqn_pairwise_potential}
\end{equation}
is used for all particle interactions. $a_{ij}$ and $r_c$ are 
constants that define the strength and range of the particle 
interaction. The magnitude of the resulting force decreases linearly 
from $\left|\mathbf F_{ij}^C(0)\right|=a_{ij}$ to 
$\left|\mathbf F_{ij}^C(r_c)\right|=0$. The $a_{ij}$'s depend on 
the type of interacting particles---and are therefore the appropriate 
location to parameterize the model. In addition, different particles 
pairs could be given different values of $r_c$ if one wants to 
effectively give particles different radii. However, in the current 
work, we choose $r_c=1$ for all bead interactions, which is the
standard in almost all DPD simulations.

For the study of information polymers and amphiphiles, individual DPD 
beads can be covalently bonded. A bond between bead $i$ and bead $j$ 
is formalized by an additional harmonic potential
\begin{equation}
    V_{ij}^s(r) = 
    \left\{
    \begin{array}{ll}
        \frac b 2 \left(r-r_b\right)^2 & 
            \mbox{ if $(i,j)$ are bonded} \\
        0 & \mbox{ otherwise}
    \end{array}
    \right.
    \label{eqn_harmonic}
\end{equation}
with bond strength $b$ and range $r_b$. In addition to that, we 
introduce a bending potential to stiffen longer polymer strands:
In a chain $i-j-k$ of interconnected polymer beads, the angle
$\theta_j$ formed by the two bonds of the central bead $j$ induces an 
additional harmonic potential 
\begin{equation}
  V_{ijk}^{\theta}(\mathbf \theta_j) = 
  \frac {\mbox{\scriptsize 1}} {\mbox{\scriptsize 2}} 
  c_{ijk} \left(\theta_j-\theta_{eq}\right)^2 \mbox{ ,}
    \label{eqn_stiffness}
\end{equation}
where $\theta_{eq}$ is the equilibrium angle and $c_{ijk}$ denotes 
the strength of the bending potential. 

The dissipative force ${\mathbf F}_{ij}^D$ is a function of the 
relative velocity of the two particles. It models the viscous damping 
of the fluid. The friction coefficient $\eta$ in eq.
(\ref{eqn_dissipative}) scales the strength of this force and 
$\omega^D$ is a distance weighting function not 
determined by the general formalism.

The random force, ${\mathbf F}_{ij}^R$ accounts for thermal effects.
It is scaled by a strength parameter $\sigma$ and a second weighting 
function $\omega^R$. $\xi_{ij}$ is a Gaussian distributed random 
variable with $\left<\xi_{ij}(t)\right>=0$, 
\(
    \left<\xi_{ij}(t)\xi_{kl}(t')\right> = 
    (\delta_{ik}\delta_{jl} + \delta_{il}\delta_{jk})\delta(t-t')
\)
and $\xi_{ij} = \xi_{ji}$.

In order to reproduce the right thermodynamic behavior, the DPD 
formalism must satisfy the fluctuation dissipation theorem. As a 
consequence, the equilibrium state will obey Maxwell-Boltzmann 
statistics and therefore allows the derivation of thermodynamic 
properties. As shown by Espa\~nol and Warren \citep{Esp:1995}, DPD 
satisfies the fluctuation dissipation theorem if and only if the 
weighting functions $\omega^D$ and $\omega^R$ obey the relation
\begin{equation}
    \omega^D = (\omega^R)^2\mbox{.}
    \label{eqn_therm_consistency}
\end{equation}
In agreement with the DPD standard, we set
\begin{equation}
    \omega^D(r) = (\omega^R(r))^2 = 
        \left[ 2(1-\frac r {r_c}) \right]^2\mbox{.}
\end{equation}
If relation (\ref{eqn_therm_consistency}) is fulfilled, 
$\mathbf F_{ij}^D + \mathbf F_{ij}^R$ acts like a thermostat to 
regulate the temperature of the system and the equilibrium 
temperature $k_bT$ is given by
\begin{equation}
    k_bT = \frac{\sigma^2}{2\eta}\mbox{.}
    \label{eqn_temperature}
\end{equation}
where $k_b$ denotes the Boltzmann constant. In molecular dynamics 
simulations, a variety of thermostats have been explored, but only the 
DPD-thermostat is guaranteed to conserve linear and angular momenta 
of the particles and thus flow properties of the fluid (because all 
involved forces are central: $\mathbf F_{ij} = -\mathbf F_{ji}$). It 
is therefore the only thermostat that allows the study of transport 
processes \citep{Tro:2003}.

In agreement with the DPD standard, we use $r_c$ and $k_bT$ as our 
units of length and energy. All particles have unit mass $m_i=1$. 
From equation (\ref{eqn_dynamics}) we can derive the unit of time as 
$\tau=r_c\sqrt{m/{k_bT}}$. We will give an estimate of the order of
magnitude of the physical length in section \ref{sec_results}.

\subsection{Incorporation of chemical reactions}
We extended the DPD formalism to account for chemical reactions. The
way that chemical reactions are implemented in our model is taken from 
\citet{Ono:2001}, where Brownian Dynamics is extended with the same 
algorithm.

Chemical reactions in our system occur between two reactants and fall 
into two different classes:
\begin{eqnarray*}
    \mbox{transformation: } \mathbf X  
        & \longrightarrow & 
        \mathbf Y\\
    \mbox{polymerization: } \mathbf X + \mathbf Y
        & \longrightarrow & 
        \mathbf{XY} \\
\end{eqnarray*}
Each reaction has a given rate for spontaneous occurrence $k_s$.

The spontaneous reaction rate can be enhanced by the presence of 
nearby catalysts. The catalytic effect decreases linearly with
increasing distance to the reactant up to a cutoff distance $r_{cat}$
beyond which it is zero. For simplicity, the effect of several 
catalysts is modeled as a superposition. Thus, the overall reaction 
rate is given as
\begin{equation}
    k = k_s + \sum_{\mathbf C} f_{cat}(r_{\mathbf C})
\end{equation}
with
\begin{equation}
    f_{cat} = \left\{
    \begin{array}{cl}
    k_{cat}\left(1-\frac{r_{\mathbf C}}{r_{cat}}\right) &
        \mbox{ if } r_{\mathbf C} < r_{cat} \\
    0 & \mbox{else}
    \end{array}
    \right.
\end{equation}
In these equations, the sum runs over all catalyst beads, with
$r_{\mathbf C}$ denoting the distance to the first reactant, $r_{cat}$ 
the maximal catalytic range, and $k_{cat}$ is the catalytic rate.
Polymerization has the further restriction that the distance between
the reactants must be less than a maximal reaction range $R$.
To deduce probabilities from the reaction rates, we used an 
agent-based like algorithm that is given in appendix
\ref{sec_algorithm}.

If a reaction occurs, we change the particle types of the reactants
from $\mathbf X$ to $\mathbf Y$ and/or establish or remove a bond 
between the reactants, depending on the type of reaction. Particle 
positions and momenta are conserved. 

We also introduced particle exchange into the model to mimic the 
supply of chemicals into the system, which drive it out of its 
equilibrium. Within a given region, particles of a certain class can 
be exchanged with a given probability to drive certain processes. Note 
that total particle number is kept constant. Likewise in chemical
reactions, we conserve positions and momenta when exchanging particles. 

\subsection{Components of the minimal protocell model}
\label{sec_components}
We model the protocell with the following components:
water, surfactant precursor, surfactant, sensitizer, information 
templates, and information oligomers and their precursors. 
Water ($\mathbf W$) and sensitizer ($\mathbf Z$) are single DPD 
particles. Surfactants are modeled as amphiphilic dimers: one 
hydrophilic head ($\mathbf H$) and one hydrophobic tail particle 
($\mathbf T$) connected by a covalent bond. Precursor surfactants are 
dimers of two hydrophobic particles ($\mathbf T-\mathbf T$).
Interaction parameters (as multiples of $k_bT$) for the water and 
amphiphiles have been taken from \cite{Gro:2000} (where surfactants 
are modeled as dimers as well):

\begin{center}
\begin{tabular}{c|ccc}
$a_{ij}$  &  $\mathbf W$ & $\mathbf H$ & $\mathbf T$ \\
\hline
$\mathbf W$ & 25 & 15 & 80 \\
$\mathbf H$ & 15 & 35 & 80 \\
$\mathbf T$ & 80 & 80 & 15 \\
\end{tabular}
\end{center}

Bond parameters are $b=150 k_bT$ and $r_b=0.5r_c$. These parameter 
values were originally used to analyze polymer surfactant
interactions. Later, the phase diagram for varying surfactant 
concentrations was analyzed \citep{Yua:2002}.

In order to keep the number of different parameters as low as 
possible, we express further interactions with the same parameters 
as the ones above: sensitizer beads are hydrophobic. Thus, their 
interaction parameters are equal to those for surfactant tails: 
$a_{\mathbf Z j}=a_{\mathbf T j}$.

\subsubsection*{Genes}

The gene is modeled as a strand of covalently bound monomers 
($\mathbf A$ and $\mathbf B$) with hydrophobic anchors ($\mathbf T$) 
attached to it. We assume the gene is similar to a peptide nucleic 
acid (PNA) decorated with lipophilic side chains to the backbone. 
The reason why we are utilizing PNA and not DNA or RNA is because 
we want to have a non-charged backbone for the gene molecule to 
enhance its lipophilic properties. For details, see \cite{Ras:2003}. 
We note that the use of PNA decorated with lipophilic side chains in 
conjunction with an amphiphilic surface layer will cause the genetic 
material to have a behavior that is quite different from that of DNA 
or RNA in water. In particular, it is not at all clear that the two 
complementary macromolecules locally will lie in a common plane when 
hybridized with each other. Thus we investigated a number of possible 
different orientations. 

By numbering the monomers within each strand, we introduce an 
orientation of the molecule that mimics the orientation of the 
actual peptide bond given by its \textit{C-} and \textit{N-termini}. 
This allows us to define the following vectors for each gene monomer 
bead: $\mathbf u_i$ is a unit vector pointing from the previous 
monomer towards the current one. For the first monomer in the strand
$\mathbf u_i=0$. Likewise, $\mathbf v_i$ is a unit vector pointing 
towards the next monomer in the strand (or $0$ for the last monomer). 
$\mathbf z_i$ is a unit vector pointing from the actual monomer 
towards its anchor bead. To obtain the association of PNA to the 
micellar surface, the molecule is modeled as interconnected 
amphiphiles. For the hydrophobic anchors, we use the same bead type
$\mathbf T$ as used for the surfactants and precursors, while 
nucleotide beads share the interaction parameters of the hydrophiles:
$a_{\mathbf A j} = a_{\mathbf B j} = a_{\mathbf H j}$.
We need to introduce additional interactions that describe the 
affinity of complementary gene monomers. Due to the rather complex 
combination of hydrogen bond formation and cooperative and $\pi$ 
stacking between real gene monomers, we cannot expect the 
complementary monomer bead forces to be as simple as the bead-bead 
interactions introduced in the former section. We now implement and
test several alternative representations of such base affinities as 
discussed below.

\paragraph{undirected attraction:}
\label{sec_attr_undir}
The obvious extension of $\mathbf F^C_{ij}$ to include attractive
interactions is a combination of attractive and repulsive components.
Thus, in the first representation, we replace 
$\mathbf F^C_{\mathbf A \mathbf B}(\mathbf r)$ by the stepwise linear
function
\begin{equation}
  \mathbf F^{C_1}_{\mathbf A \mathbf B}(\mathbf r) =
  \mathbf F^C_{\mathbf A \mathbf B}(\mathbf r) + \left\{
  \begin{array}{cl}
  a_2\left(r_{c_2} - r\right) \mathbf n & \mbox{ if } r<r_{c_2}\\
  0 & \mbox{ else}
  \end{array}
  \right.
\end{equation}
with $r_{c_2}>r_c$ and $a_2<0$.
Different attraction strengths $a_2$ will be used and compared in
later computer simulations (section \ref{sec_hybridization}).
To compensate strong attractions for small values of $r$, 
we will vary the repulsion strength  $a_1=a_{\mathbf{AB}}$ 
accordingly. Note that another generalization of 
$\mathbf F^{C_1}_{\mathbf A \mathbf B}$ compared to 
$\mathbf F^{C}_{\mathbf{AB}}$ is the change in the interaction range 
which, in addition to the standard $r_c$ dependence, now also depends
on the actual pair $(\mathbf A,\mathbf B)$ through $r_{c_2}$.

\paragraph{directed ``radial'' attraction:}
\label{sec_attr_rad}
In the real gene system, hybridization is partly due to the formation 
of H-bonds between the complementary nucleotides. H-bonds share 
features with covalent bonds, which are better characterized by 
directed rather than radial interactions. Hence, in the second 
representation, we introduce directed attraction parallel to the 
$\mathbf A-\mathbf T$ and $\mathbf B-\mathbf T$ axes, respectively.
Here, we replace $\mathbf F^C_{\mathbf A \mathbf B}$ by
\begin{equation}
  \mathbf F^{C_2}_{\mathbf A \mathbf B}(\mathbf r) = 
  \mathbf F^C_{\mathbf A \mathbf B}(\mathbf r) \:+\:
  \left\{
  \begin{array}{cl}
    a_2 (r_{c_2}-r) \left(\mathbf z\cdot\mathbf r\right) 
    \mathbf n & \mbox{ if } r<r_{c_2} \\
    0 & \mbox{ else}
  \label{eqn_attract_radial}
  \end{array}
  \right.
\end{equation}
with the above definitions for $\mathbf r$, $\mathbf z$, and 
$\mathbf n$.
Again, different attraction coefficients $a_2$ will be compared in
the later simulations. The value $a_1=a_{\mathbf{AB}}$, on the other
hand, can be held fixed because the attraction vanishes when $r$ 
approaches 0. We set $a_1 = 35k_bT = a_{\mathbf{AA}}=a_{\mathbf{BB}}$
We call this interaction ``radial'', because the strongest
attraction will be radial towards the center of the micelle, once the 
PNA strand is attached to the surface of the micelle.

\paragraph{directed ``tangential'' attraction:}
\label{sec_attr_tang}
The third representation is similar to the second, except that
attraction is now perpendicular to the backbone and to the 
$\mathbf{AT}$ (or $\mathbf{BT}$) axis. The force is attractive towards
one side of the PNA and repulsive towards the other---hence, it is the
only implementation that catches the directionality of the molecule:
\begin{equation}
    \mathbf F^{C_3}_{\mathbf{AB}}(\mathbf r) =
    \mathbf F^C_{\mathbf{AB}}(\mathbf r) + \left\{
        \begin{array}{cl}
            a_2 (r_{c_2}-r)
                \left( \frac
                    {\mathbf (u + \mathbf v) \times \mathbf z}
                    {\left|\mathbf(u+\mathbf v)\times\mathbf z\right|}
                \cdot \mathbf r \right) 
                \mathbf n & \mbox{ if } r<r_{c_2} \\
            0 & \mbox{ else}
        \end{array}
    \right.
\end{equation}
This force is expected to be strongest tangential to the surface of
the micelle. As in the last case, we will vary $a_2$, but keep $a_1$ 
fixed at a value of $35k_bT$.

\begin{figure}[hbt]
    \centering
    \includegraphics[width=.8\columnwidth]{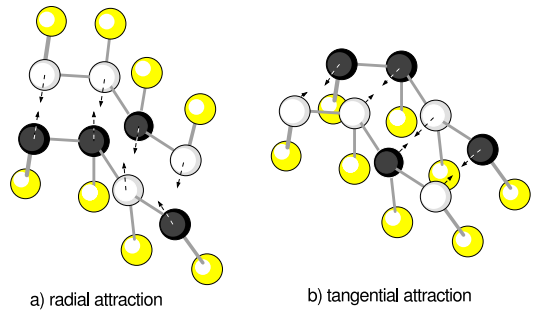}
    \caption{
        Hybridization complexes for radial (a) and tangential (b)
        attraction between complementary bases. Bases are shown
        as black and white beads, hydrophobic anchors in yellow.
        Arrows denote the direction of strongest attraction.
    }
    \label{fig_pna_forces}
\end{figure}

\par
Covalent bonds within PNA strands have a bond strength of $b=150k_bT$ 
with an ideal bond length $r_b=0.5r_c$ for bonds between nucleotides 
and anchors, and $r_b=0.75r_c$ for bonds between the nucleotides 
themselves. In addition, we introduce stiffness (eq.
\ref{eqn_stiffness}) within the PNA strand: angles of interconnected 
strands prefer to be stretched out ($\theta_0=180^o$, 
$c_{ijk}=10k_bT$). With the stiffness we model folding restrictions 
of the peptide bond, as well as $\pi$-$\pi$ electron stacking of 
nearby nucleotides. This affects only the PNA chain, not the bonded 
hydrophobic anchors, as they do not experience any bending 
potential. Table \ref{tab_interact} summarizes the chosen set of 
parameters.
\begin{table}[hbt]
\centering
\begin{tabular}{c|ccccccc}
$a_{ij}$  &  $\mathbf W$ & $\mathbf H$ & $\mathbf T$ & 
                                   $\mathbf A$ & $\mathbf B$ & $\mathbf Z$ \\
\hline
$\mathbf W$ & 25 & 15 & 80 & 15 & 15 & 80 \\
$\mathbf H$ & 15 & 35 & 80 & 35 & 35 & 80 \\
$\mathbf T$ & 80 & 80 & 15 & 80 & 80 & 15 \\
$\mathbf A$ & 15 & 35 & 80 & 35 &(*) & 80 \\
$\mathbf A$ & 15 & 35 & 80 &(*) & 35 & 80 \\
$\mathbf Z$ & 80 & 80 & 15 & 80 & 80 & 15
\end{tabular}
\caption{
Interaction strength $a_{ij}$ (as multiples of $k_bT$) for all bead 
types defined in the model. The force (*) between complementary 
nucleotides $\mathbf A$ and $\mathbf B$ has attractive parts and 
cannot be expressed by a single interaction parameter 
$a_{\mathbf A\mathbf B}$. Three different force fields have been 
considered for such interactions. See the text for details.
}
\label{tab_interact}
\end{table}

\subsubsection*{Reactions}

For the above listed components we introduce the following chemical 
reactions:

First, we define a reaction that transforms the oil-like precursor
surfactants into actual surfactants. In the real chemical 
implementation of the protocell, the precursors are fatty acid 
esters. The ester bond of the precursor surfactant breaks thereby 
producing a fatty acid---the surfactant---and some small aromatic 
molecule---which is considered waste. Disregarding the production of
the waste, we model this reaction by the scheme
\begin{equation}
    \mathbf{TT} + \mathbf Z \longrightarrow \mathbf{HT} + \mathbf Z
    \label{eqn_metabolism}
\end{equation}
which reflects, that both parts of the ester are hydrophobic, while
the resulting surfactant is an amphiphile. For simplicity, the
spontaneous reaction rate is set to $0\tau^{-1}$. The sensitizer acts
as a catalyst with a catalytic radius of $1.0 r_c$. In our simulation,
the catalytic rate of the sensitizer can be turned on 
($k_{cat}=1.0\tau^{-1}$) and off ($0\tau^{-1}$) interactively 
by a switch. This mimics the photo-activity of the sensitizer.

Second, we introduce reactions to form covalent bonds between the
terminal monomers of pairs of oligomers.
\begin{eqnarray}
    \mathbf A + \mathbf B & \longrightarrow & \mathbf{AB} \nonumber \\
    \mathbf A + \mathbf A & \longrightarrow & \mathbf{AA} \\
    \mathbf B + \mathbf B & \longrightarrow & \mathbf{BB} \nonumber
\end{eqnarray}
These syntheses are only applied to the terminal monomers in the PNA 
strands and involve no catalysts. The maximal range is $0.75r_c$, the 
maximal reaction rate is $k_{max}=0.1\tau^{-1}$. The actual reaction 
rate between monomers $i$ and $j$ further depends on the orientation 
of the ligating strands: we set
\begin{equation}
  k_{ij} = \frac 1 2 k_{max} 
      \left(
        \frac{\mathbf u_i+\mathbf v_i} 2 \cdot 
        \frac{\mathbf u_j+\mathbf v_j} 2
        + 1
      \right)
\end{equation}
This formulation also prevents covalent bonds between 
complementary strands (which are anti-parallel, and thus, have an 
effective $k$ close to zero).

\section{Results}
\label{sec_results}
We use the model discussed above to study various aspects of the life 
cycle of the Los Alamos Bug as depicted in figure \ref{fig_cartoon}. 
In particular, our simulations address the spontaneous self-assembly 
of protocells (Fig. \ref{fig_cartoon}, frames 1\&2), the incorporation 
of resources (frames 2\&3), the metabolic growth of the 
protocell (frames 4\&5), template reproduction, and finally 
fission into two daughter cells (frames 5\&6). We will 
further analyze some of the catalytic coupling processes explained in 
the introduction.

All simulations are performed in three-dimensional space with periodic 
boundaries. We set $\sigma$ to 3 and $\eta$ to 4.5, which leads to an 
equilibrium temperature of $1 k_bT$. A total bead density 
$\rho=3.0r_c^{-3}$ is used for all simulations. System size and 
number of iterations is noted for each individual simulation run.
We integrate equation 
(\ref{eqn_dynamics}) numerically with the DPD variant of the leapfrog 
Verlet integrator discussed in \cite{Gro:1997} with $\lambda=0.5$ 
and a numerical step width of $\Delta t=0.04\tau$.

\subsection{Self-assembly of micelles}
\label{sec_micelles}
\begin{figure}[htb]
  \centering
  \includegraphics[width=\columnwidth]{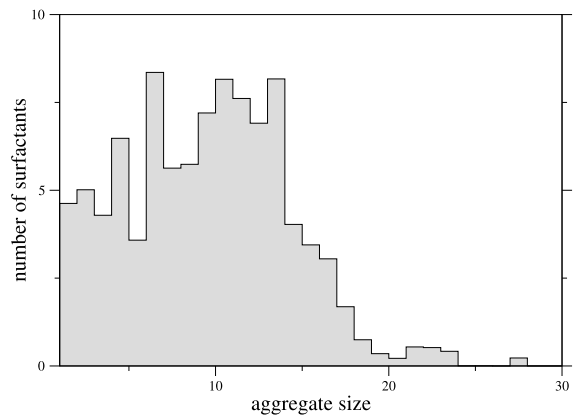}
  \caption{
    Micellar size distribution for a system containing $2.9$ water
    beads and $0.05$ surfactant dimers per unit cube. To obtain the
    aggregate size histograms from a system state, every two 
    surfactants whose $\mathbf T$-beads are separated by less than 
    $r_c$ are considered to belong to the same aggregate. 20000
    systems states of an equilibrated system 
    ($200\tau < t < 1000\tau$) are averaged in the shown distribution.
  }
  \label{fig_avgdistr}
\end{figure}

We initialize a cubic box of size $(12.5r_c)^3$ randomly with $2.9$ 
water beads and $0.05$ surfactant dimers per unit volume, or 5664
water beads and 98 dimers in the box. Simulations 
are performed for $0\tau < t < 1000\tau$ with the interaction 
parameters summarized in Table \ref{tab_interact} and the model
parameters given in the introduction to this section. We observe the 
formation of spherical micelles with aggregation numbers up to about
20, with a peak around 12. This is shown in figure \ref{fig_avgdistr},
where once the system had reached an equilibrium state, we followed
its behavior. For each time step we recorded the number of aggregates
of a particular aggregation number and hence the total number of
surfactants in the aggregates of that size. The average of this result
over the number of time steps was than histogrammed. We also 
observe
a continuous exchange of surfactants with the bulk phase. As a result 
of these associations and dissociations, we find a number of free 
monomers and sub-micellar aggregates in the bulk phase. These
observations qualitatively fit theoretical and experimental results
\citep[see e.g.][]{Eva:1999}.

Although we do not intend to model specific chemicals, we can roughly
estimate the order of magnitude for the physical length scale of our
simulation, using a procedure proposed by \citet{Gro:2001}. Our
calculation is based on sodium alkanesulfates as these are well
studied surfactants with properties similar to the fatty acids used in
the real chemical implementation. Table \ref{tab_surfactants} lists
the critical micelle concentration (CMC), i.e. the minimal
concentration at which micelles spontaneously form. The table also
gives the mean aggregation number and the volume of these
molecules. 
\begin{table*}
\begin{tabular}{|l|ccc|cccc|}
\hline
surfactant & CMC                & aggregation  & surfactant vol. & $N_m$ & $r_c$       & surfactant conc. & predicted \\
           & in mol/l           & number       & in $\mbox{\AA}^3$ & & in $\mbox{\AA}$ & in mol/l         & micellization ratio \\
\hline
$NaC_6H_{13}SO_4$    & $0.42$             & $17\pm6$     & 278 & 4.625 & 7.467 & 0.201 & $1\cdot10^{-5}$   \\
$NaC_7H_{15}SO_4$    & $0.22$             & $22\pm10$    & 305 & 5.075 & 7.701 & 0.183 & $2.5\cdot10^{-3}$ \\
$NaC_8H_{17}SO_4$    & $0.13$             & $27$         & 332 & 5.525 & 7.923 & 0.168 & 0.2   \\
$NaC_9H_{19}SO_4$    & $6.0\cdot 10^{-2}$ & $33$         & 359 & 5.975 & 8.132 & 0.156 & 0.6   \\
$NaC_{11}H_{23}SO_4$ & $1.6\cdot 10^{-2}$ & $52$         & 413 & 6.875 & 8.521 & 0.135 & 0.935 \\
$NaC_{12}H_{25}SO_4$ & $8.2\cdot 10^{-3}$ & $64\pm13$    & 440 & 7.325 & 8.703 & 0.127 & ~1    \\
$NaC_{14}H_{29}SO_4$ & $2.1\cdot 10^{-3}$ & $80\pm16.5$  & 494 & 8.225 & 9.046 & 0.113 & ~1    \\
\hline
\end{tabular}
\caption{
    Data for sodium alkanesulfate surfactants with varying tail 
    length. For each surfactant, CMC and mean aggregation number are 
    listed from \cite{Ani:1976}. The molecular volume is estimated
    from the number $n$ of carbon atoms using the formula 
    $V=27(n+1)\mbox{\AA}^3$ \citep{Eva:1999} plus a constant
    $88.51\mbox{\AA}^3$ for the sulfate group (whose value is derived
    from the molecular mass ($98.08g/mol$) and density ($1.84g/cm^3$) 
    of sulfuric acid). The coarse graining parameter $N_m$, the 
    physical length scale $r_c$, and the total surfactant 
    concentration are the interpretation of model parameters in case
    that the model dimer represents the respective surfactant. 
    Finally, the fraction of micellized surfactant is the prediction 
    of the closed association model for the respective surfactant and 
    the calculated concentration \citep{Eva:1999}.
}
\label{tab_surfactants}
\end{table*}
Under the simplifying assumption that all DPD beads have equal 
effective volume, we can derive the molecular volume of a single DPD 
bead and -- knowing the molecular volume of water
($V_\text{H$_2$O}=30\mbox{\AA}^3$) -- we get the so-called
\textit{coarse graining parameter}
\begin{equation}
    N_m = \frac{\frac 1 2 V_\text{surf}}{V_\text{H$_2$O}}
\end{equation}
that tells us, how many water molecules are represented by a single 
DPD bead. The average number of DPD water beads per unit cube is 
$\rho$, each one of them representing $N_m$ molecules. Therefore, 
the physical length scale $r_c$ resolves to
\begin{equation}
    r_c \equiv (\rho N_m V_\text{H$_2$O})^{1/3} \mbox{.}
\end{equation}
We will work with solutions that are quite dilute and hence dominated
by water. Noting that a liter of water has $1000/18=55.56$ moles of
water in it, while a volume of $r_c^3$ has $\rho N_m$ molecules of
water in it, we find that a concentration of 1 particle/$r_c^3$ yields
a unit of concentration as
\begin{equation}
    1 r_c^{-3} \equiv 55.56 mol/\rho N_m \mbox{.}
\end{equation}

With these estimations, we find that the lipid concentration in the 
above simulation represents between $0.11$ and 0.20 $mol/l$. It is 
somewhat arguable to estimate the concentration of free lipids in the 
bulk phase, because our simulations do not yield a sharp distinction
between free lipids---i.e. submicellar aggregates---and proper 
micelles. Assuming that the most reasonable choice for such a 
distinction is the first minimum in the micellar size distribution at 
aggregates of size 5 or less, from figure \ref{fig_avgdistr} we get 
an average of 22.9 free surfactants in the bulk phase out of 98
lipids in the total volume, i.e. 76.6\% of the surfactant is
micellized and the free lipid concentration lies between $0.03$ and 
0.05 $mol/l$. 
Knowing the physical surfactant concentration, we can compare this 
finding to the prediction of the closed association model 
\citep{Eva:1999}. According to this model, surfactants are either in
bulk phase ($S$) or in micelles of aggregation number $N$ ($S_N$).
With the pseudo-chemical reaction $NS \rightleftharpoons S_N$ and the
condition that
$
\left.\frac{d[S]}{d[S]_\text{total}}\right|_\text{CMC}
= \left.\frac{dN[S_N]}{d[S]_\text{total}}\right|_\text{CMC}
= 0.5
$,
one can calculate the fraction of micellized surfactant for any 
total surfactant concentration $[S]_\text{total}=[S] + N[S_N]$. The
respective ratio $N[S_N]/[S]_\text{total}$ is also given in 
table \ref{tab_surfactants}.

We find that our model best matches 
the aggregation numbers of short chain surfactants 
($NaC_6H_{13}SO_4$), while our micellization ratios more closely match
the predictions for the somewhat longer chains ($NaC_9H_{19}SO_4$).
Although our model representation of surfactants as dimers is rather
simplistic, we find a reasonable match (at least in the order of
magnitude) between experiment, simulation, and theory. It should be
noticed that the micellization parameters for fatty acids, which are
the container surfactants of choice in the Los Alamos Bug, are
qualitatively similar to the listed sodium alkanesulfate surfactant
parameters, which are the most well studied surfactants in the 
scientific community. Given the easy availability of relevant 
parameters for alkanesulfate surfactant parameters and the level of
coarse graining in our DPD model we can safely use these experimental
data to calibrate our simulation. It is 
conceivable that closer matches might be found by changing interaction
parameters or the representation of surfactants. We have however
decided to stick to the standard parameter set in order to get 
comparable results to earlier DPD simulations 
\citep{Gro:2000,Yua:2002,Fel:2006}. 

Next, we analyzed a ternary mixture of water, surfactant, and oil. In 
the system described above, we exchanged an additional $0.1$ water 
beads per unit volume by $0.05$ hydrophobic oil dimers 
($\mathbf T-\mathbf T$),
which represent the lipid precursors of the Los Alamos Bug. Starting 
from a random initial condition, the system forms loaded micelles: 
the precursors aggregate into a core in the interior of the individual
micelles because of their high degree of hydrophobicity. This core is 
coated by surfactants, which shield it from water. We observe a 
stabilizing effect from the hydrophobic core: the rate of monomer 
dissociation from the aggregates decreases by a factor of 4 to 5. 
Dissociation of oil dimers does not happen during the simulations. 
Over the simulated time span ($0\tau < t < 1000\tau$), these loaded 
micelles constantly fused to form bigger aggregates. At $t=250\tau$, 
the system is composed of five micelles with aggregation numbers 12, 
13, 16, 24, and 32, where the aggregation number just counts the 
surfactants in an aggregate and not any of the precursors or other 
components. At $t=500\tau$ we find four micelles (with sizes 16, 24, 
25, 32) and finally, for $t=1000\tau$, the system consisted of only
two micelles with aggregation numbers 43 and 53. It remains unclear, 
whether this was the equilibrium solution, or whether the two micelles
would finally fuse to form a single aggregate. It is known that any 
given mixture of surfactants and oil in water results in some 
equilibrium aggregate structure, some useful and some less useful as 
a protocellular container substrate, see e.g. the recent summary 
discussion in \cite{Cas:2006}.


In general, the addition of hydrophobic precursors allows aggregates 
to grow far beyond their micellar aggregation number, while at the 
same time, monomer dissociations from the assembly falls by a factor 
of four or more. This is consistent with simulation results from 
earlier studies of a similar surfactant-precursor-water system 
\citep{Fel:2006}. However, a more systematic DPD
investigation is necessary to address the dynamics, stability, and
size distribution issue in this context.

\subsection{Self-assembly of the protocell}
In this section, we study the self-assembly of protocells. We 
initialize a cubic box of size $(7.5r_c)^3$ with 1212 water particles,
21 surfactant dimers, 4 sensitizer particles and one PNA strand that
is four nucleotides in length. All other simulation parameters are as 
before. Using these numbers, we achieve the same overall particle 
density and the same surfactant concentration as in the previous 
section.

\begin{figure}[bt]
  \centering
  \includegraphics[width=.32\columnwidth]{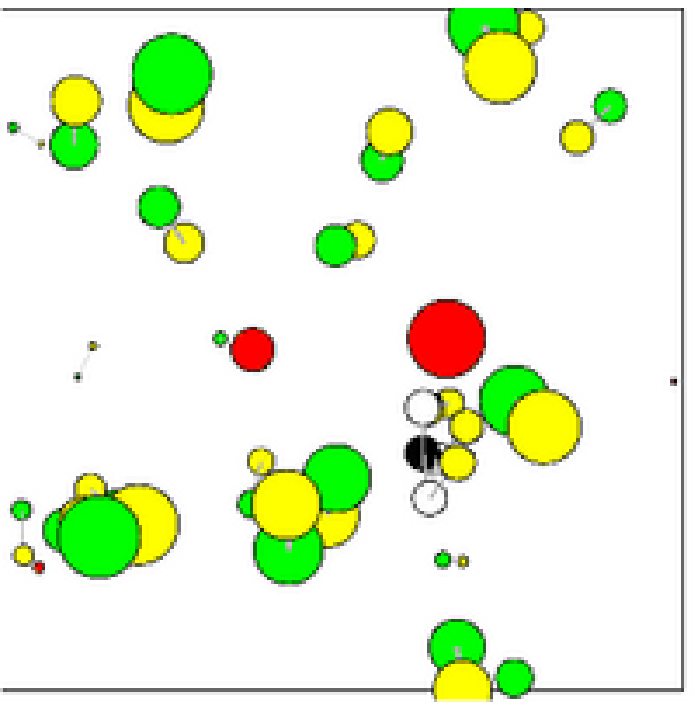}
  \includegraphics[width=.32\columnwidth]{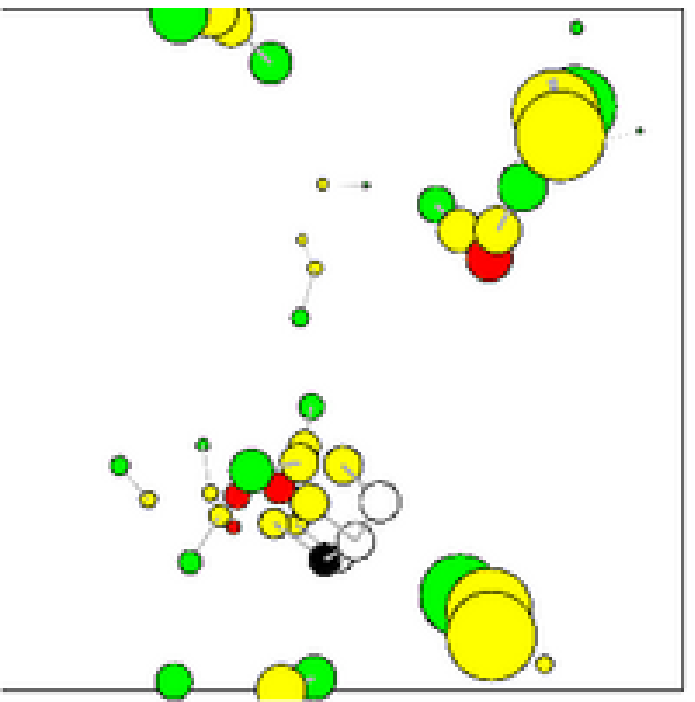}
  \includegraphics[width=.32\columnwidth]{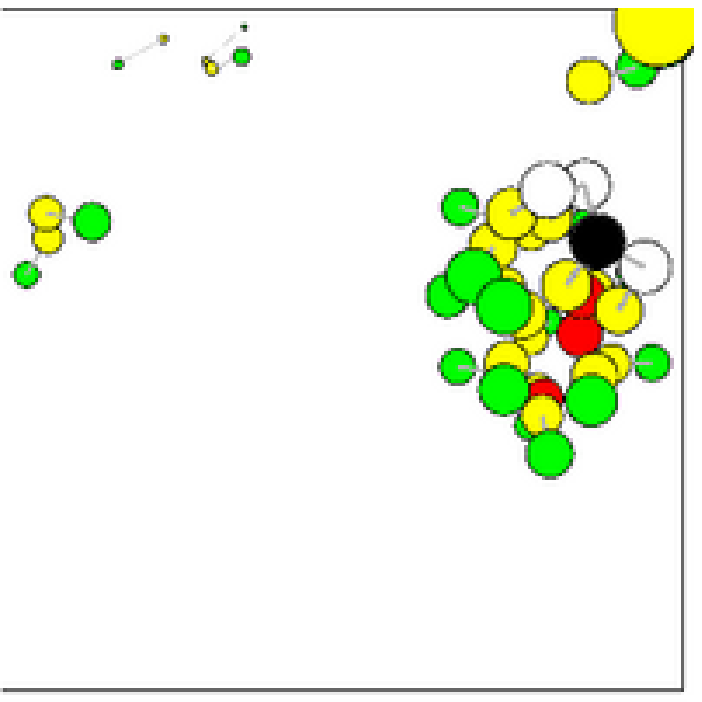}
  \caption{
    Self-assembly of the protocell from a random initial condition. 
    The diagrams show the state of the system at times a) $t=0\tau$,
    b) $t=4\tau$, and c) $t=10\tau$. 
    Surfactants are shown in green (head bead) and yellow (tail), 
    the sensitizers in red, the PNA backbone in yellow and the PNA 
    monomers in black and white. 
  }
  \label{fig_assembly}
\end{figure}

Starting from an arbitrary initial condition, we observe the 
spontaneous formation of a protocell, i.e. a micelle that is loaded 
with sensitizer and which has PNA attached to its surface and whose 
nucleotides are exposed to the aqueous phase (see figure 
\ref{fig_assembly}). Aggregation happens within a remarkably short 
period: after only 10 time units, we already find complete protocells. 
The lipid aggregation number of this micelle is around 14 with few 
associations and dissociations of monomers. The slight increase in 
aggregation number along with a decrease of monomer dissociations is 
most probably due to the stabilizing effect of the additional 
sensitizers.

\subsection{Replication of the Container}
\label{seq_container}
The dynamics of a surfactant-precursor-water system similar to the one 
under consideration has been studied in detail in \cite{Fel:2006}. 
Considering precursor and surfactant kinetics, the formerly analyzed 
system differs from the one discussed here in that i) the catalytic
role of sensitizers is performed by the surfactants themselves, and 
ii) the metabolic turnover is not regulated by turning the light on 
and off, but instead only follows chemical mass kinetics. Using
simulations based on classical lattice gas methods, 
\citeauthor{Cov:1996} in \citeyear{Cov:1996} reproduced the 
micellar self-replication experiments of \citet{Bac:1992}. 
In \citeyear{May:1998} and \citeyear{May:2000} \citeauthor{May:2000}
developed an extended lattice polymer approach for explicitly 
including polymers and chemical reactions similar to the current DPD 
approach and they were also able to reproduce the experimental 
findings by Luisi's group \citep{Bac:1992}. The purpose of this 
section is to show that the reported dynamics also hold for the 
metabolic reaction scheme of the Los Alamos Bug.

A system of size $(10r_c)^3$ is initialized with a micelle consisting 
of 15 surfactants and loaded with 4 sensitizer beads in its interior.
Model parameters are given in the beginning of this section.
In a single spherical region of radius $2r_c$ located away
from the micelle, pairs of water particles are replaced by surfactant 
dimer precursors with an overall exchange rate of 
$\approx 2.5\times 10^{-3}$ precursors per time unit.

Because of their hydrophobic nature, the precursor molecules tend to 
agglomerate into oil-like droplets. The diffusion of such droplets 
becomes progressively slower the bigger they are. This initiates a 
positive feedback: the bigger the droplets, the more slowly they 
diffuse out of the exchange 
region. The slower they diffuse, the more likely they are to 
accumulate additional precursors before they diffuse out of the 
exchange volume. By varying the volume of the exchange region and/or 
the rate of exchange, one can set the mean size of the precursor 
droplets that are formed. Due to the positive feedback, the effect 
will not be linear with either the exchange region size or the
exchange rate. 

Since we do not want the non-continuous exchange events to disturb 
the systems dynamics too much, we restrict particle exchange to a 
region of $2.0r_c$ (3\% of the total system volume). By varying the 
exchange rate used to introduce precursors, we find that 
$5.0\times 10^{-5}$ is close to the optimum for which droplets of 
precursor molecules are provided at a reasonable rate, yet are still 
small enough to diffuse at a reasonable speed. With these values, the 
precursor droplets consisted of 5 dimers on average. Once in the 
vicinity of a micelle, the droplets are immediately absorbed. 

\begin{figure}[bt]
    \centering
    \includegraphics[width=.32\columnwidth]{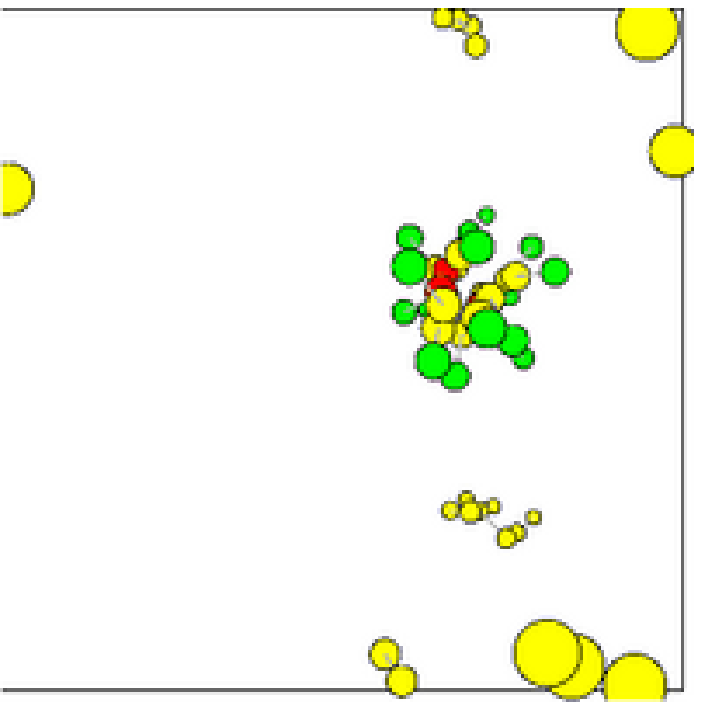}
    \includegraphics[width=.32\columnwidth]{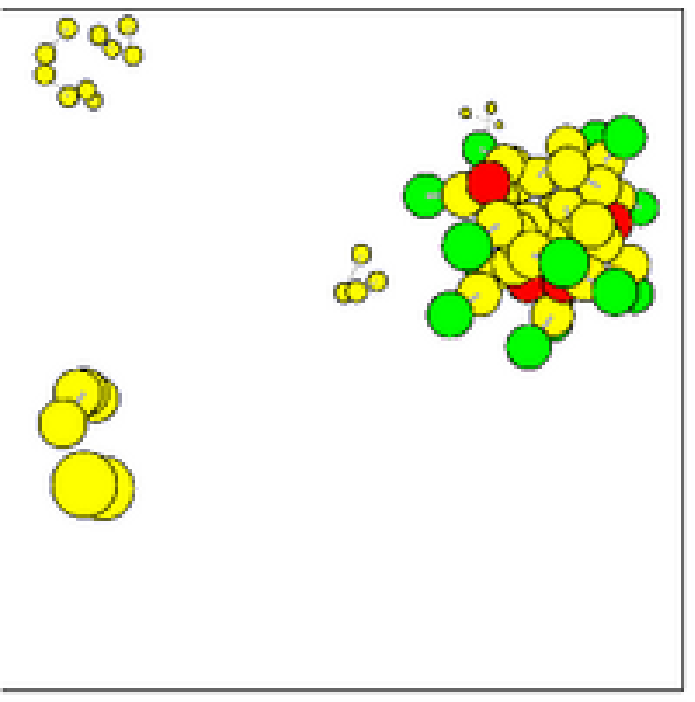}
    \includegraphics[width=.32\columnwidth]{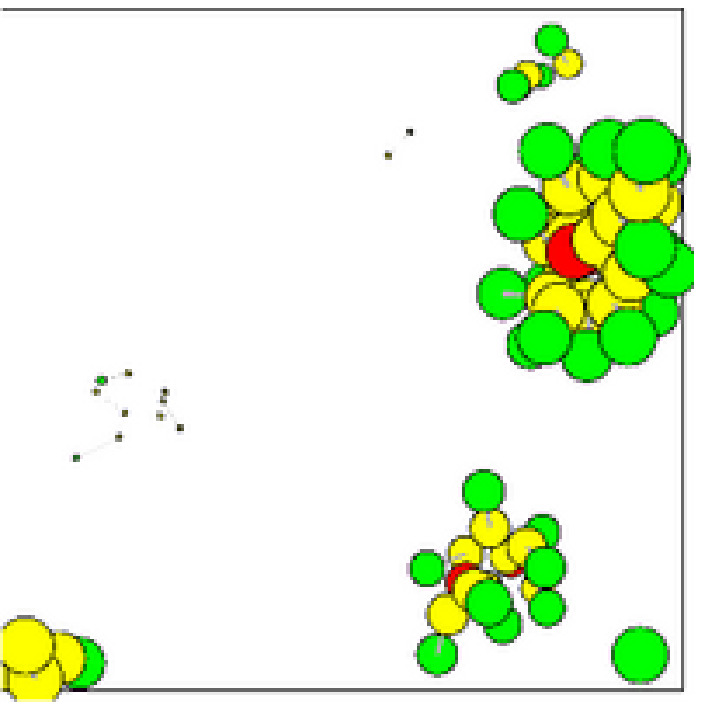}
    \caption{
        Replication dynamics of the container: precursors are fed 
        into the system far from the micelle at the (periodically 
        reflected) edge of the system space. They form droplets in 
        the aqueous phase, which are absorbed by the protocells as a 
        whole. Protocells grow by incorporation of precursors. After 
        a critical amount of precursor is transformed into 
        surfactant, the assembly loses its stability and splits in 
        two daughter cells (right frame).
    }
    \label{fig_growth}
\end{figure}

When the micelle absorbs 15 precursor molecules into its interior, we 
stop supplying additional precursors and trigger the catalytic 
activity of the sensitizer by turning on the light. During the 
metabolic turnover, the micelle grows in amphiphile number, while 
losing few, if any, amphiphiles due to the stabilizing effect of the 
remaining precursors as was discussed previously. It responds to the 
changing surfactant to precursor ratio by changing its shape from 
spherical to rod-like. The elongation continues until nearly all the 
precursors are metabolized. At some moment, the elongated aggregate 
becomes unstable and divides into two daughter cells (see figure 
\ref{fig_growth}). With the parameters used, overall precursor 
turnover and fission takes place in approximately 20 time units
(i.e., 500 time steps). 

We compared the above findings to simulations of an unregulated 
system, where the precursor supply and catalytic rate are not 
triggered, but instead held constant over the whole simulated time 
span. The objective behind this simulation was to find whether the 
system might feature inherent self-regulation: as the precursor forms 
droplets in the bulk phase, their incorporation into the micelle 
occurs in spurts rather than continuously. If the introduction rate 
of precursors into the system is locally fast enough to allow larger 
droplets to form (especially due to the positive feedback effect), a 
larger number of precursors can simultaneously enter the protocell. 
Then if the metabolic turnover rate is sufficiently fast, the 
turnover of the large number of precursors might be sufficient to 
trigger container division rather than having a slow but continual 
loss of newly formed amphiphiles.

To investigate this possibility, we performed simulation runs for a 
system of size $(10r_c)^3$ initialized with a micelle of 15 
surfactants and 4 sensitizer beads. Other model parameters are the
same as given in the beginning of this section. Precursors were 
supplied by the same mechanism and rate as before. We observed the 
incorporation of droplets between 3 and 9 precursor dimers in size. 
As the transformation of precursors happened significantly faster 
than the precursor supply, nearly each droplet was transformed 
separately. When only few precursors were absorbed at once (i.e. a 
small droplet), the micelle responded by rejecting several 
surfactants into the bulk phase. Such loose surfactants then formed 
sub-micellar aggregates or attached to precursor droplets when 
present. However, when the incorporated droplet was big enough, the 
outcome of the metabolic turnover was a proper cell division. A 
micelle that consisted of 15 surfactants and 4 sensitizers, for 
example, split in two after the absorption and turnover of 8 
precursors. The fission products were two micelles, one with 14 
surfactants and 3 sensitizers and the other with 9 surfactants and 
1 sensitizer.

This result suggests that the explicit regulation of the metabolic 
turnover by light bursts might not be necessary to obtain the 
replication cycle of the container as a similar regulation can be 
obtained by a careful regulation of the provided precursor droplet 
sizes. Light control might, however, still serve as a convenient 
mechanism to synchronize container and genome replication if they 
occur on separate time scales.

\subsection{Replication of the genome}
In our experience, the most difficult component of the protocell to 
model with DPD methods is the genome and its behavior. Furthermore, 
the DPD hybridization process seems more illdefined than the 
ligation process, which is why our discussion of the replication of 
the genome is divided in two consecutive steps: hybridization and 
ligation. 
Please recall that hybridization denotes the alignment of short PNA 
oligomer sections along the template PNA strand and ``hydrogen'' 
bonding to it, while ligation---or polymerization---is the reaction 
that turns aligned oligomers into an actual (complementary) copy of 
the template.

\subsubsection{Hybridization}
\label{sec_hybridization}
\begin{figure}
    \centering
    \includegraphics[width=.4\columnwidth]{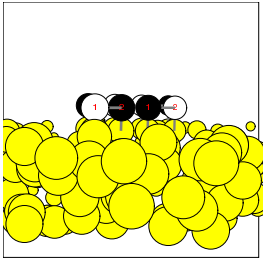}
    \caption{
        Initial setup of the hybridization simulations. The system is
        initialized with an oil-layer that mimics the oil-water 
        interface of a two-phase system. A four-mer template and two 
        complementary dimers are placed at the interface so that they 
        form a hybridization complex. The association time of such 
        hybridization complexes is measured for different PNA 
        implementations and attraction forces.
    \label{fig_initialhybrid}
    }
\end{figure}

Replication of the genome essentially depends on the stability of the
hybridized complex: it can only occur if the double strands are stable 
for a time long enough for all the needed oligomers to diffuse to and 
align with the template. It should be noted that if more than 2
oligomers are involved, the joining of additional oligomers and their
polymerization can occur sequentially so the unpolymerized templates
need not all be simultaneously attached. As will be shown further 
below, once some polymerization has occurred, that section will be 
more stable in hybridized form. We studied the stability of the 
hybridization with the following simulation: A system of size 
$(5.5r_c)^3$ was initialized with an oil layer that is meant to mimic 
a two phase system (single beads of type $\mathbf T$ are confined to 
lie below a plane above which the water is located). 
The overall particle density is $\rho=3r_c^{-3}$, as in the earlier 
experiments. in order to make the hybridization process as 
simple as possible. As we shall see later, aggregate surfactant dimers
tend to 
tangle with the gene anchors, which both slows down the hybridization 
process and makes it less accurate. A four-monomer long PNA template 
was placed at the oil-water-interface with its anchors pointing down
toward the oil and its bases pointing up towards the aqueous phase. 
A pair of 2-nucleotide long complementary oligomers was placed 
at a distance of $0.5r_c$ from this strand at a location/orientation
for proper hybridization. The location/orientation was 
varied to match the different hybridization cases studied. In the 
case of directed radial attraction, this meant that all the beads of 
the complementary PNA molecules are outside the interface plane, 
with their hydrophobic anchors pointing away from the hybridization 
site. In contrast, in the case of tangential attraction, both the 
template and the oligomers span the interface region as shown in 
figure \ref{fig_initialhybrid}.

In the system modeled, we only had two different types of monomers 
($\mathbf A$, $\mathbf B$) with $\mathbf A$ and $\mathbf B$ being 
complementary to each other, but not self-complementary. All different 
4-mer templates excluding symmetric configurations were used (e.g. 
$\mathbf{AAAA}$, $\mathbf{AAAB}$, $\mathbf{AABB}$, $\mathbf{ABAB}$,
and $\mathbf{ABBA}$) and for each different template only the proper 
complementary dimer oligomers were used. The different 4-mer 
configurations can differentially hinder the ability of the 
complementary bases to slide along the template.

During the
simulations, the distances between all four complementary base pairs 
were measured at every time step. When one of these distances exceeded
$1.5r_c$ (the maximal interaction range for complementary bases), the 
PNA strands were considered to be dehybridized. The time it took for 
the double strands to dehybridize---i.e. the association time of the 
hybridized complex---serves as a measure of the stability of that 
state. After a maximum of $t=100\tau$, simulations were truncated and 
the hybridization was considered to be stable.

\begin{figure}[hbt]
  \centering
  \includegraphics[height=\columnwidth,angle=270]
    {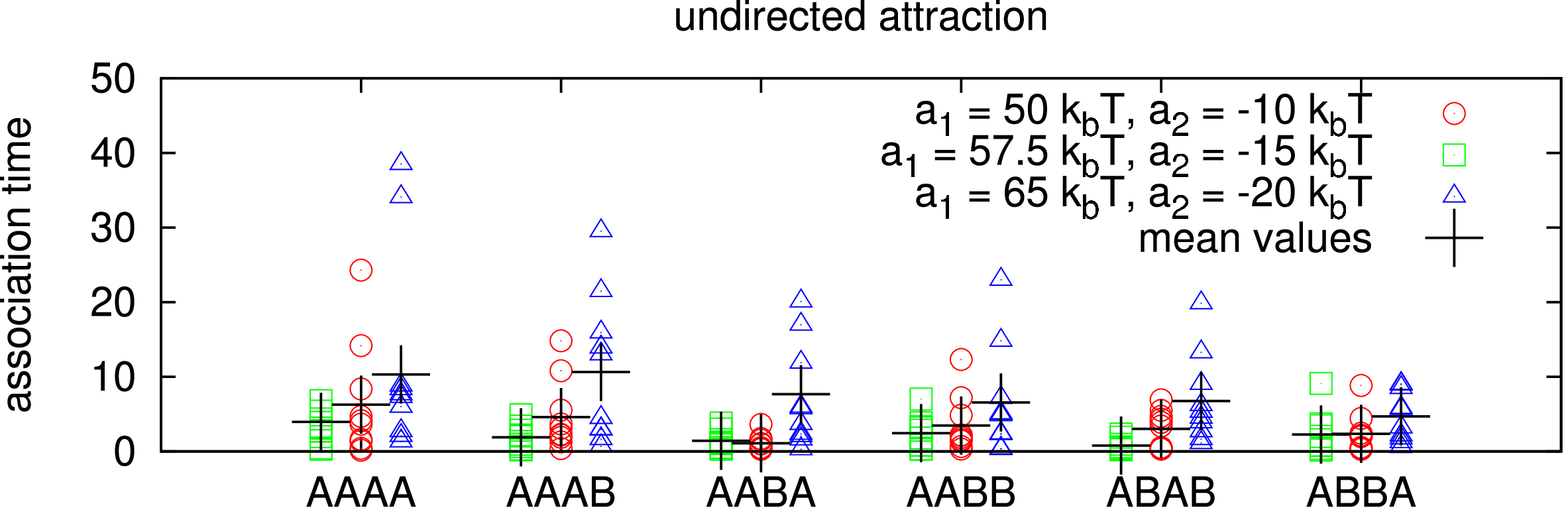}
  \vspace{.5em}
  
  \includegraphics[height=\columnwidth,angle=270]
    {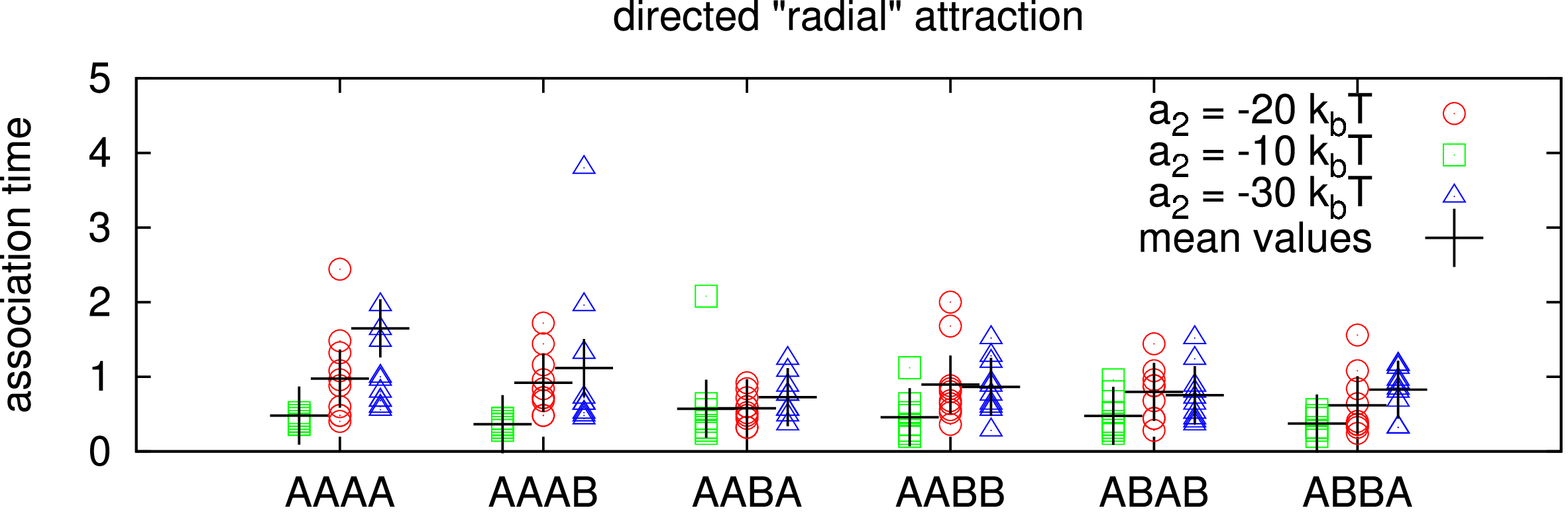}
  \vspace{.5em}
  
  \includegraphics[height=\columnwidth,angle=270]
    {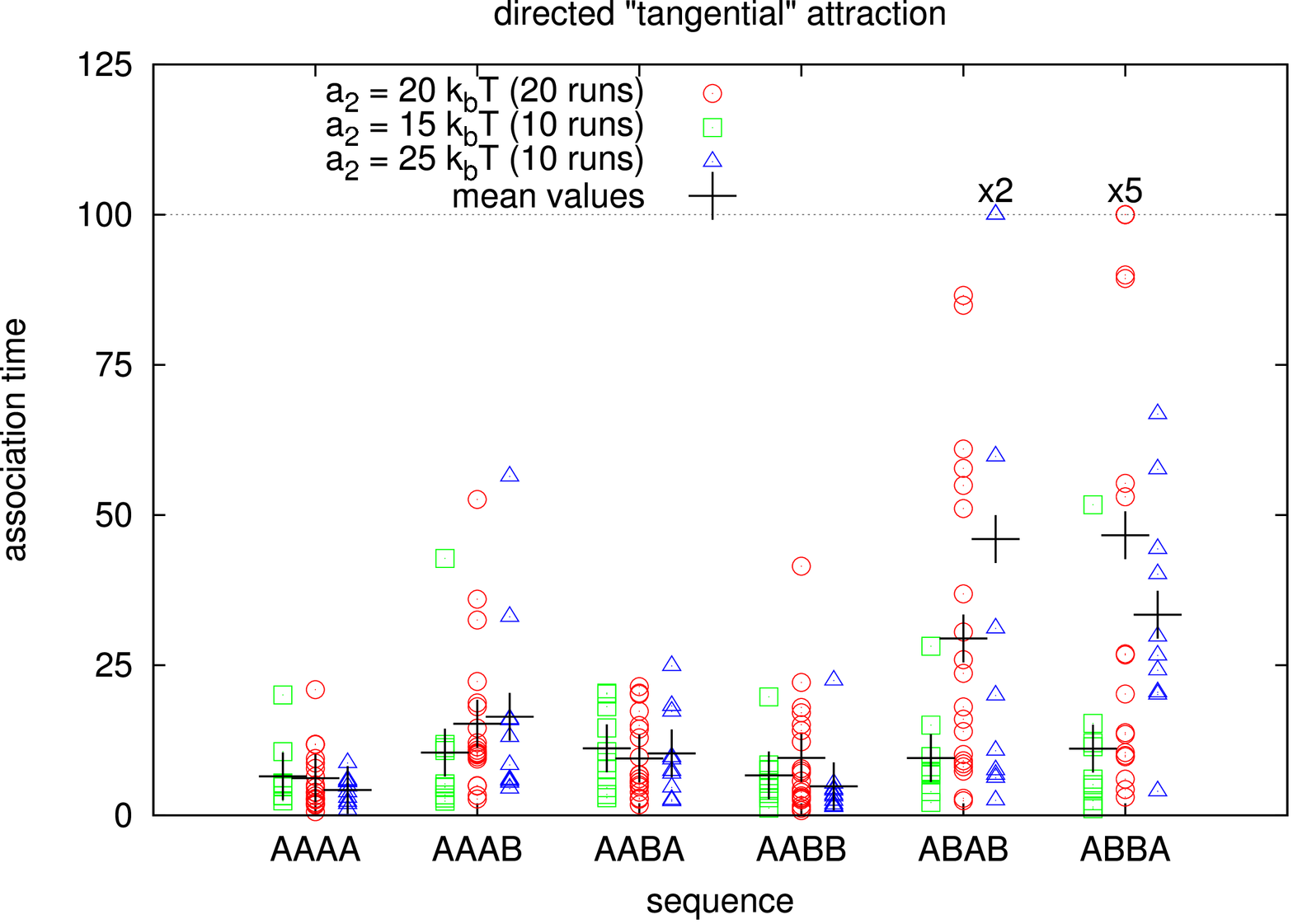}
  \caption{
    The association times (i.e. the time until the initially 
    hybridized complex becomes dehybridized) for different PNA 
    template sequences of length four using a) undirected, b) radial, 
    and c) tangential attraction. For each implementation, three 
    different attraction strengths are compared, as given in the 
    legend for each figure. $a_1$ denotes the coefficient of the 
    repulsive part, $a_2$ the coefficient of the attractive part of 
    the interaction force. In the case of directed attraction 
    (b and c) $a_1$ was set to $35k_bT$ independent of the 
    respective value of $a_2$. In c), the plotted averages are 
    minimal values for the actual averages, as simulations were 
    truncated at $t=100\tau$. If runs were truncated, the multipliers 
    above that run designate how often this was done. 
  }
  \label{fig_hybridization}
\end{figure}

For the three different representations of PNA hybridization
(see sections \ref{sec_components} \textit{Genes}, cases a,b, and c), 
we performed simulations for all possible combinations of four bases 
excluding symmetrical 
combinations. Strengths for attractive forces were set with respect 
to the repulsive force parameter $a_{\mathbf A \mathbf B}$ so that 
complementary bases attracted each other but did not overlap by more 
than $0.6r_c$. The association times were measured using 10 to 20 
runs for each combination. 
Results are shown in figure \ref{fig_hybridization}.

\paragraph{undirected attraction:}
In the case of undirected attraction, we found mean association times 
between $2.12\tau$ for $a_1=50k_bT$, $a_2=-10k_bT$, and $7.76\tau$  
for $a_1=65k_bT$, $a_2=-20k_bT$. For strong 
attractions, association times tended to increase with the number of 
equal (preferably nearby) nucleotides in the template 
($\mathbf{AAAA}$ is the most, while $\mathbf{ABBA}$ is the least 
stable sequence). However, these differences were rather small.

\paragraph{directed radial attraction:}
For directed radial attraction, the mean association times ranged from
$0.45\tau$ for $a_2=-10k_bT$ to $0.98\tau$ for $a_2=-30k_bT$ 
($a_1=a_{\mathbf{AB}}=35k_bT$ for all cases) without any significant 
variation for different sequences. For most simulation runs, it took 
only a few time steps for the initial complex to dehybridize. 
The reason for the poor nature of the hybridization of the PNA for 
the radial attraction is quite obvious: due to the amphiphilic 
character of PNA, the strands will arrange so that nucleotides point 
towards water and the anchors towards oil. Thus, the attraction is 
directed perpendicular to the oil-water interface and into the aqueous
phase where the oligomers do not want to reside. Because of the dot 
product in equation (\ref{eqn_attract_radial}), the attraction between 
two PNA molecules on the interface is marginal and the association 
time is essentially a matter of diffusion.

\paragraph{directed tangential attraction:} 
In contrast to the other tested situations, in the case of directed 
tangential attraction, one can see significant differences in the 
association time of the initial hybridized complexes, provided the 
attraction is strong enough: for gene sequences with pairs of equal 
bases at terminal positions (e.g. $\mathbf{AAAA}$ and 
$\mathbf{AABB}$), hybridization is usually less stable than for 
sequences without equal bases at terminal positions ($\mathbf{ABBA}$ 
and $\mathbf{ABAB}$). The association time of sequences with only one 
such dimer lies between the values of the above two situations.
Examination of the simulations reveal the cause of this trend: a
continuous group of two or more equal monomers, one of which is a
terminal position of the template allows the attached dimer
to slide along the template strand without a strong penalty in 
potential energy, and eventually protrude beyond the end of the 
template. In this misaligned configuration, the dimer can easily 
distort from the parallel alignment, thereby reducing the overall 
attraction to the template, until it finally disassociates from the
complex. Distinct bases at terminal positions, on the contrary, 
prevent this sliding along and then off of the template, thereby 
significantly stabilizing the hybridized state.

\begin{figure}
  \centering
  \includegraphics[height=\columnwidth,angle=270]{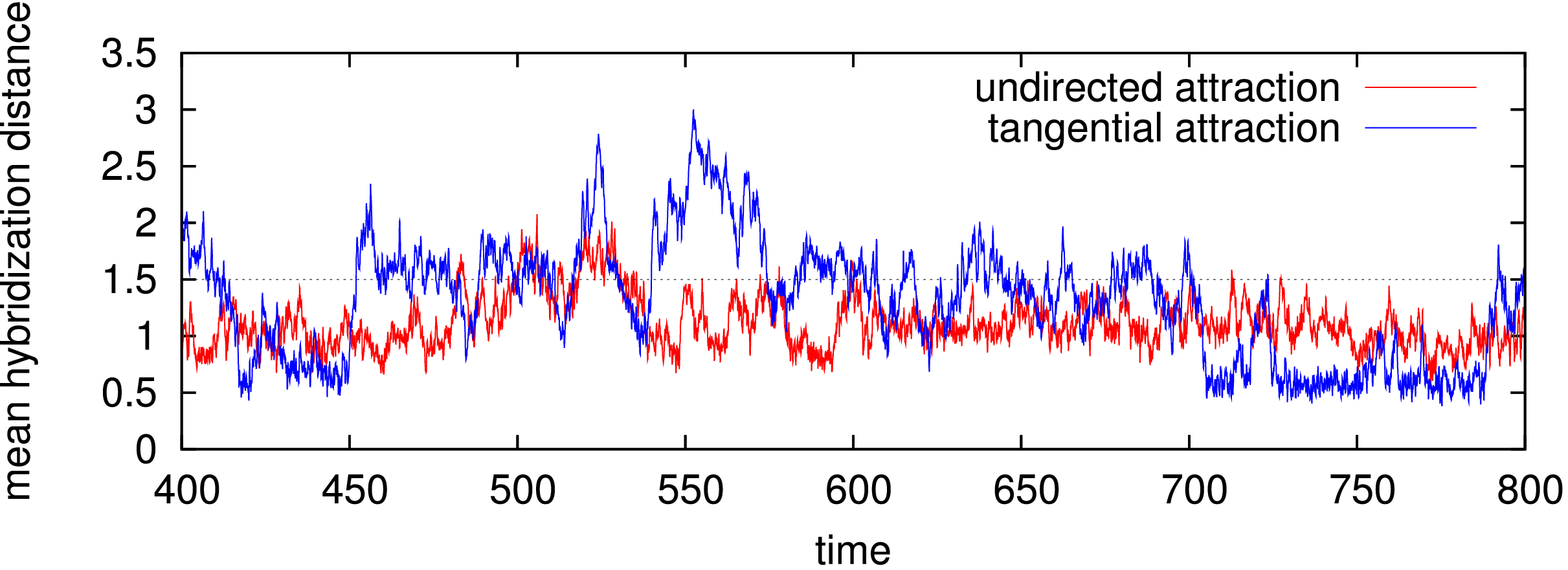}
  \includegraphics[height=\columnwidth,angle=270]{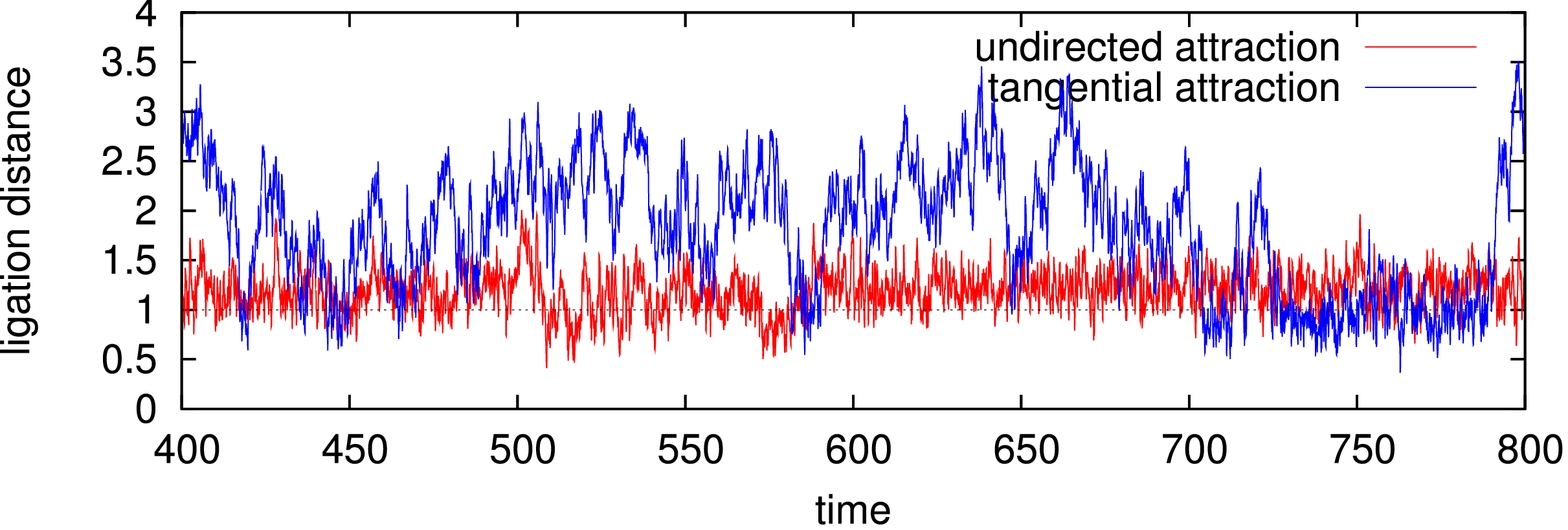}
  \caption{
    Mean hybridization (upper panel) and ligation distance (lower 
    panel) for the PNA templates (and corresponding oligomers) 
    $\mathbf{AAAA}$ using undirected attraction with 
    $a_1=65k_bT$, $a_2=-20k_bT$ (red) and for $\mathbf{ABBA}$ using 
    tangential attraction with $a_1=35k_vT, a_2=40k_bT$ (blue).     
    By hybridization distance, we mean the average distance between 
    complementary nucleotides, by ligation distance, we mean the 
    minimal distance between two terminal nucleotides that are 
    supposed to polymerize.
    The maximal values of the various distances are limited by the 
    small size of the box.
  }
  \label{fig_hybridization_dist}
\end{figure}

For the more promising PNA 
implementations---undirected and tangential attraction---we further 
measured the mean distance between complementary bases (hybridization 
distance) and the distance between those bases in the oligomers that 
are supposed to polymerize (ligation distance). We performed these 
measurements using the sequence $\mathbf{AAAA}$ for the undirected, 
and $\mathbf{ABBA}$ for the tangential attractions (interaction 
parameters are given in the caption of figure 
\ref{fig_hybridization_dist}). Simulations are performed for 
$0\tau\le t\le 1000\tau$. The resulting time series are shown in 
figure \ref{fig_hybridization_dist}.

In the case of tangential hybridization one finds two alternating 
domains in the hybridization distance time series: (i) when oligomers 
are aligned to the template, the mean hybridization distance is 
around $1.04 r_c$ with only small fluctuations and an average ligation
distance of $1.01r_c$ (e.g. $430\tau\le t\le 450\tau$ and
$700\tau\le t\le780\tau$ in figure \ref{fig_hybridization_dist}). In 
between such periods, (ii) oligomers dissociate from the template, and 
diffuse over the interface, which is indicated by the large variance 
in hybridization distance.

Undirected attraction, in contrast, yields hybridization distances 
around $1.07r_c$ with significant continual fluctuations and a mean 
ligation distance of $1.158r_c$. One cannot observe the ``locking''
of the hybridized state that is apparent for the tangential 
attraction: although the oligomers preferably stay in the vicinity of 
the template, they are not forced into any particular orientation. 
Investigation of simulation states reveals that oligomers align 
along different sites of the template or even cross the template 
strand. Thus, although it appears from a quick look at figure 
\ref{fig_hybridization_dist} that the undirected attraction performs
better on average, it is only during the ``locked in'' period that the
desired reactions occur. We can therefore conclude that only the 
implementation of PNA 
using tangential attraction is able to generate a proper 
hybridization and base recognition approximation.

\begin{figure}
  \centering
  \includegraphics[height=\columnwidth,angle=270]
    {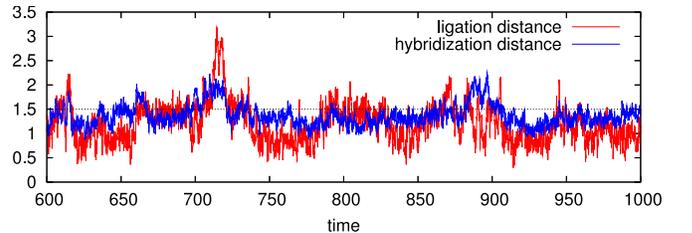}
  \caption{Hybridization and ligation distances of PNA template and
    complementary oligomers in water. For PNA, tangential directed 
    attraction with $a_2=40k_bT$ has been used. The nucleotide 
    sequence is $\mathbf{ABBA}$.
  }
  \label{fig_hybrid_water}
\end{figure}

It is assumed that the PNA replication is catalyzed by the oil-water 
or surfactant-water interface.  This is because: (i) lipophilic PNA 
concentrates at the oil-water interface and thus obtains a much 
higher local concentration there than in water; (ii) that the 
interface contains a lower water concentration than the bulk phase; 
(iii) that the interface might directly act as a catalyst for the 
amide bond formation; and (iv) that the PNA is more spread out 
(linear) when attached to the interface. To test the geometric part 
of this hypothesis, we also performed simulations of hybridization in 
pure water. We randomly initialized a box of size $(5.5r_c)^3$
with water, PNA template ($\mathbf{ABBA}$) and complementary 
oligomers using directed tangential forces (the overall bead density
was $\rho=3r_c^{-3}$). Simulations were performed for 
$0\tau\le t\le1000\tau$. Hybridization and ligation distances are 
plotted in figure \ref{fig_hybrid_water}.
The mean hybridization distance in this scenario is $1.41r_c$ (which 
is close to the maximum radius $r_{c_2}$ at which attraction of 
complementary nucleotides still exists) with a standard deviation of 
$0.34r_c$. Moreover, there is no clear separation between hybridized 
and dehybridized states. In contrast to the scenario for the oil-water 
interface, the oligomers never completely dissociate from the
template. However, the oligomers are not properly hybridized either. 
Instead, the template and complementary strands mainly attract each 
other due to the hydrophobic interactions between the tail beads of 
these strands rather than forces between their bases. Inspection 
of the simulated states shows that oligomers are seldom aligned 
parallel to the template. The overall structure has more resemblance 
to that of a micelle with geometries defined by the amphiphilic 
properties of the molecules, rather than a double strand defined by 
base affinities. The ligation distance has an average value of 
$1.12 r_c$ with a standard deviation of $0.39r_c$. Unfortunately, 
this is smaller than in the previous simulations. This might result 
in ligation rates higher than those on the surface. However, if we 
decide to vary the ligation probability depending on the angle 
between PNA backbones, the effective ligation rate is smaller than 
at the oil-water interface. 

Last but not least, it is notable that we cannot achieve reliable
hybridization without a stiffness potential in the PNA chain. In the 
absence of such stiffness, complementary bases within one strand tend 
to bind to each other and form sharp hairpin loops even for very 
short strands. This effectively hinders any proper hybridization 
except for very few sequences that do not offer any possibility
for loop formation (such as $\mathbf{AAAA}$).

\subsubsection{Ligation}
\begin{figure}
    \includegraphics[width=.32\columnwidth]{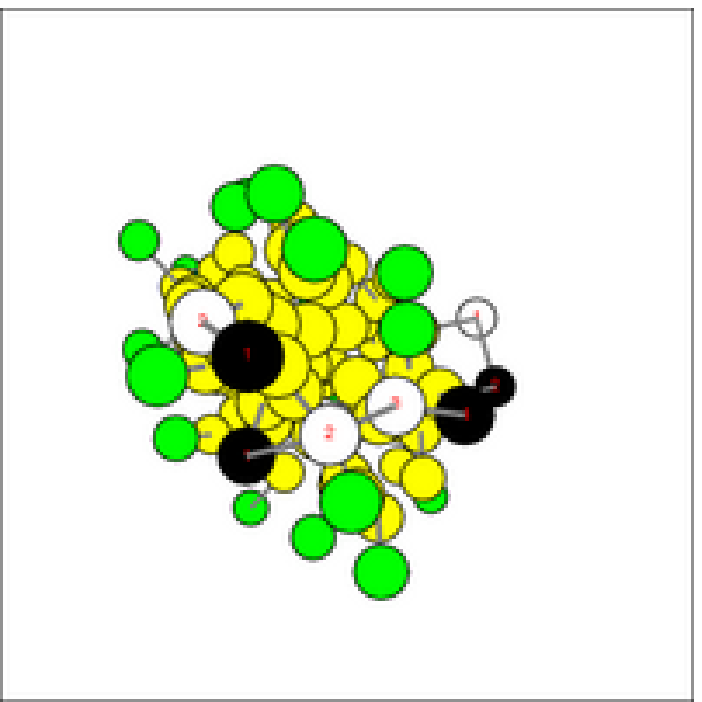}
    \includegraphics[width=.32\columnwidth]{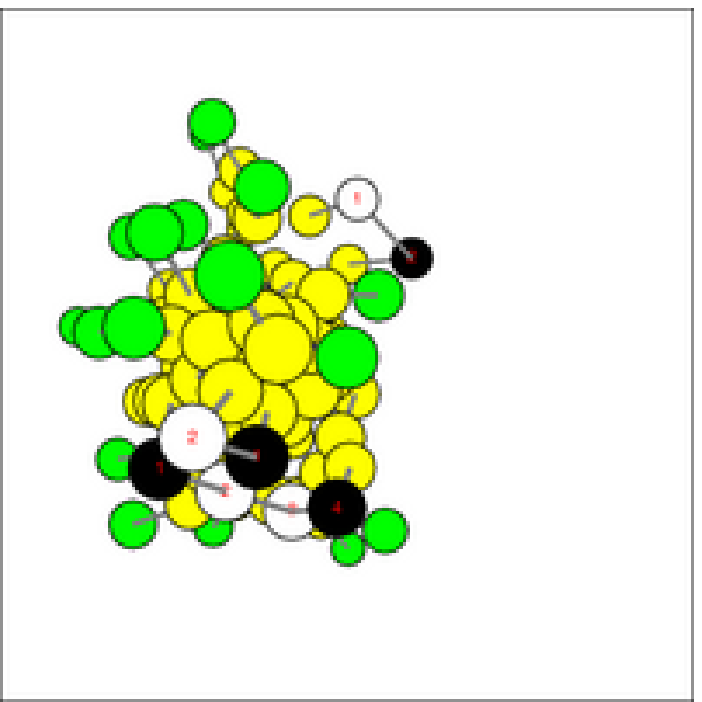}
    \includegraphics[width=.32\columnwidth]{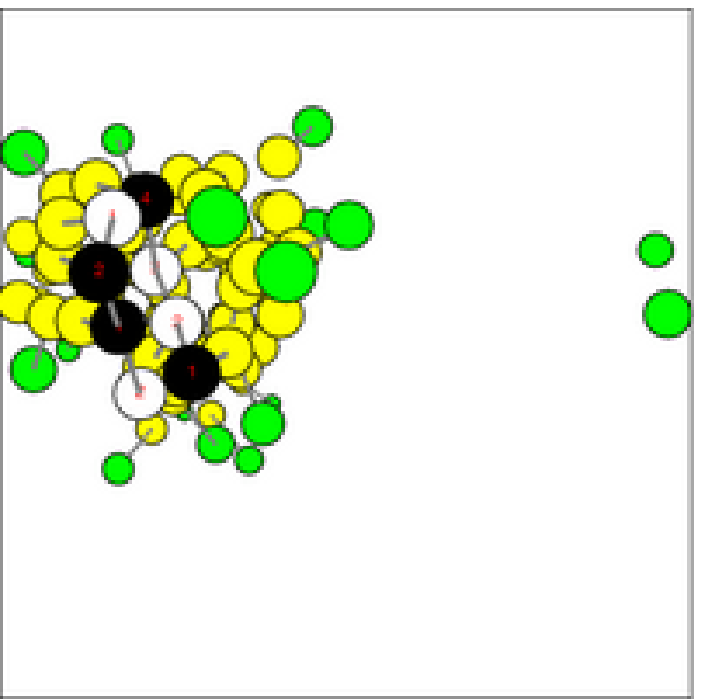}
    \caption{
        The three steps of template directed replication: a) Template
        ($\mathbf{ABBA}$) and oligomers ($\mathbf{BA}$ and 
        $\mathbf{AB}$) diffuse over the surface of the micelle, b) 
        oligomers form a hybridization complex with the template 
        strand, and c) oligomers polymerize to yield a complementary 
        copy of the template.
    }
    \label{fig_replication}
\end{figure}

To study the polymerization reaction, a four-mer template strand and
two complementary dimers are placed randomly on the
surface of a loaded micelle (20 surfactant, 20 precursors) within
a system of size $(10r_c^3)$ and total density $\rho=3.0r_c^{-3}$.
As the last section identified $\mathbf{ABBA}$ to form the most stable 
hybridization complex, we restrict polymerization experiments to this 
particular sequence using the PNA representation with tangential 
directed attraction (see figure \ref{fig_replication}). 

Of the performed simulations, 8 out of 10 generated proper template 
directed ligation, while the remaining 2 reactions occur spontaneously 
in the absence of the template strand and define the expected 
background reaction \citep{Nie:2007}. In our simulations, one of the 
two spontaneous ligation results was a correct complementary copy of 
the template strand while the other was not. Note that in our 
simulation, polymerization has not been explicitly restricted to 
happen only between C- and N-terminals, which means that both ends 
can be concatenated with any other end. When ligation is template 
directed, 6 out of 8 runs lead to correct complementary sequences, 
while the other two resulted in mispairings of the form 
$\mathbf{BABA}$. In summary, we find that correct replication is 
about $50\%$ more reliable, when directed by the template. If one 
prohibits the ligation of equal terminal beads (C-C and N-N), the 
reliability of replication is expected to further increase.

The simulations reveal that it can take a surprisingly long time for 
the oligomers to form a ligated hybridized complex with the template. 
Ligation occurs after $90\tau$ in the fastest and after $674\tau$ in 
the slowest run. The average time is estimated as $223.2\tau$.
The huge variance is due to the random walk of template and 
oligomers over the surface of the micelle. Compared to the oil-water 
interface of the previous section, oligomer motion is further slowed 
down by the head particles of the amphiphiles as well as the dimer 
structure of the aggregate building blocks.

It is worth mentioning that as expected, the hybridized complex 
is significantly more stable after the ligation has occured than 
before. None of the hybridized complexes that formed in the above 
simulations showed any sign of dissociation within 750 time units
after ligation took place. 

\subsection{Full protocell division}
\begin{figure}
    \includegraphics[width=.32\columnwidth]{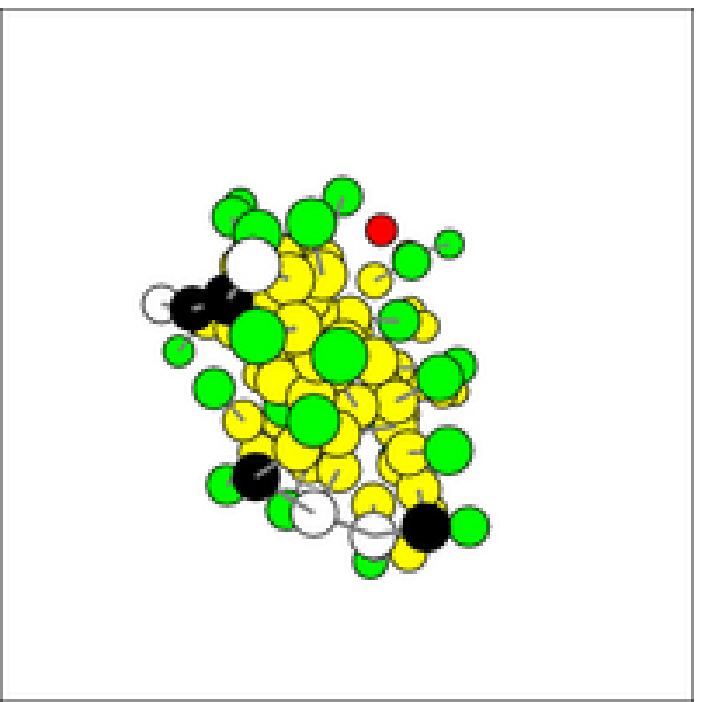}
    \includegraphics[width=.32\columnwidth]{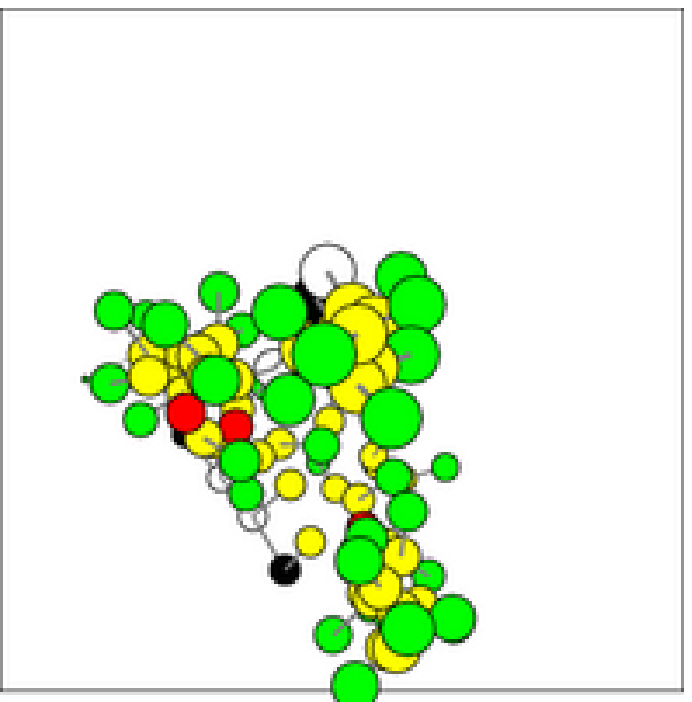}
    \includegraphics[width=.32\columnwidth]{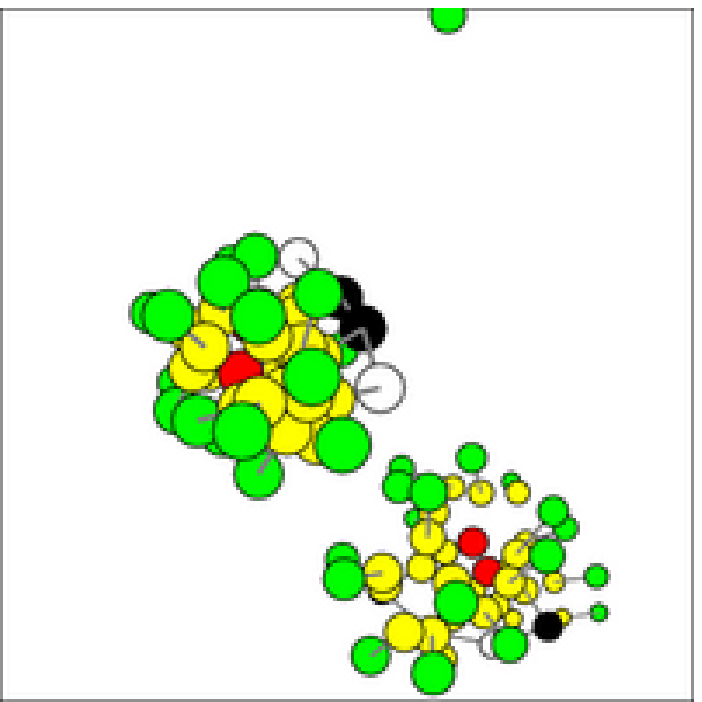}
    \caption{
        The division of the whole protocell completes the life cycle 
        of the Los Alamos Bug. A mature protocell is loaded with 
        precursor molecules, sensitizers, and two complementary PNA 
        strands. During the metabolic turnover of precursors, the 
        aggregate elongates and divides. Both PNA strands and
        sensitizer molecules tend to distribute evenly among the 
        daughter cells, when only few sensitizers are present.
    }
    \label{fig_fission}
\end{figure}

The last elemental step in the life cycle of the Los Alamos Bug is the 
fission of the grown cell into two daughter cells as shown in figure 
\ref{fig_fission}. In addition to what was discussed in 
section \ref{seq_container}, here we studied the fission of the whole 
protocell after the replication of its genome, that is, a micelle 
loaded with some lipid precursors, sensitizers and two complementary 
PNA templates. The objective  is to illuminate how templates and 
sensitizers are distributed among the daughter cells. Although not 
addressed by simulations in earlier sections, here the influence of
the number of sensitizers is also investigated. 

Proper division into two daughter 
cells requires the melting of the double stranded PNA resulting 
from genome replication, which may be achieved by a temperature cycle. 
In the DPD formalism, temperature translates into the interaction 
parameters $a_{ij}$. To study the impact of a 
temperature cycle on the whole system, one would need to exchange the 
interaction parameters between all DPD beads. For simplicity in these 
initial investigations, and in the absence of a rigorous calibration 
of our model, we chose to invoke melting by simply turning off the 
attractive hybridization interactions between the PNA bases.

We performed simulations of a system of size $(10r_c)^3$ with an 
initial protocell consisting of 20 surfactants, 20 precursors, 4 to 8 
sensitizers, and two PNA template strands randomly located on its 
surface. Otherwise, the standard parameters listed in the beginning 
of this section were used. Snapshots of the system are shown in Fig. 
\ref{fig_fission}.In all cases, metabolic turnover initiated 
the division of the aggregate at times of between 50 to 100 $\tau$
after the start of the simulation. Fission times were found to be 
longer than in the former experiments. This was because the aggregate
consisted of more particles and because the template strands 
stabilized the rod-like aggregate that precedes 
protocell division. It was observed that PNA strands were preferably 
located along the elongated part of the aggregate, rather than at the 
caps. We believe that due to the stiffness parameter (eq.
(\ref{eqn_stiffness})) of the PNA strands, the aggregate tends to 
elongate in a direction that is parallel to the PNA’s long axis.

Using only 4 sensitizers, the distribution of sensitizers and PNA 
among the daughter cells was rather diverse: in one out of 10 runs, 
all sensitizers and templates remained in one of the fission products, 
while the other consisted of only 11 surfactants. In 7 of the runs
the partition was nearly even: both sensitizers and templates were 
equally distributed among the two daughter cells, which differed in 
aggregation number by at most 3 surfactants. Last but not least, we 
also observed two runs where the other components were distributed 
equally, but one of the daughter cells contained both template 
strands. We note that although it was not observed, it might be 
possible for a template to connect two otherwise divided aggregates 
by attaching to both their surfaces. 

One might expect the equipartition of sensitizers is more likely
when their number is increased. Our simulation results, however, 
showed quite the opposite: protocells loaded with 8 sensitizers 
instead of 4 almost always responded by rejecting an average of 11 to 
12 surfactants. By doing so, the protocell was able to maintain a 
stable spherical shape even with an aggregate number of 27 
surfactants. This is due to the collective stabilizing effect of the
strongly hydrophobic core of sensitizers within the aggregate. The 
more sensitizers that are added, the more they will tend to stick 
together. 
The more they stick together, the less likely they will 
partition into different daughter cells. Thus they are better able 
to stabilize the amphiphilic dimers in the aggregate.  
For an initial protocell that holds 6 sensitizers, proper division 
can still be observed, but the results are less reliable than in the 
case of 4 sensitizers. For 6 sensitizers, equipartition of sensitizers 
was only achieved in one out of five simulations. The other runs lead 
to empty micelles or a situation where one of the daughter micelles
has only one sensitizer bead. 
Equipartition of PNA could not be achieved for the cases with either 
6 or 8 sensitizer beads.

\section{Discussion}

Because of the inherent simplifications of the aggregated DPD 
simulation technique and due to the inherent complexity of our 
protocell system, accurate predictions of neither the detailed 
kinetic nor thermodynamic properties could be expected. However, 
insights into generic issues and likely system behavior could be 
obtained by the illumination of the systemic properties of the 
proposed protocell design. In particular we were able to see how the 
global behavior emerges from the simple and well-defined properties 
of the underlying molecular ingredients. Interpolation between 
several simulation methods combined with experimental data is 
necessary to obtain predictive understanding of this protocellular 
system. Investigations based on quantum mechanics, molecular dynamics,
reaction kinetics, combined with these and other DPD studies, 
hopefully can address the quantitative prediction issues in a more 
complete manner \citep{PAsPACE}.

We found that the micellar kinetics that underlie the container 
replication are highly affected by hydrophobic molecules present in 
the solution. In the design of the Los Alamos Bug, these hydrophobic 
molecules can be the metabolic precursors and sensitizers. As these 
molecules are incorporated into the protocell, they form a core that 
stabilizes the aggregates. Such loaded micelles have a larger 
aggregation number than micelles in a pure surfactant-water system, 
and the surfactant exchange with the bulk phase is strongly decreased. 
The simulations thus suggest that a 3-component (ternary) 
surfactant-oil-water system is more suitable for yielding a suitable 
container than a two-component system based on surfactant and water 
only.

We also observed that protocells grow in spurts rather than 
continuously, even with a continuous supply of resource molecules. 
This is because the oil-like precursor molecules form droplets before 
they are absorbed by the aggregates. Furthermore, due to slower 
diffusion of larger objects, once the droplets start to form, 
volume-wise they will tend to grow ever more rapidly the larger they 
become prior to being absorbed. The spurt-like support of resources 
might be sufficient to initiate the division process of the aggregate 
if these droplets have the appropriate size. If so, the system would 
be self-regulated and no further triggering of the metabolism as with 
an external light switch would be necessary. Whether or not this 
self-regulation enables a reliable replication of the whole organism 
also depends on a number of other factors such as the rate of 
precursor supply compared to the replication rate of the genome. 
Further simulation investigations will be necessary to identify 
whether the metabolic self-regulation is sufficient when the 
precursor supply rate is not carefully balanced.


Our representations of the biopolymer that stores genomic information 
can be considered to be the crudest feature of the model. None of 
the implementations relate in detail to the actual physicochemical 
traits of the real PNA molecule. The behavior of the PNA molecule 
with hydrophobic 
side chains in our protocell is also found to be quite different from 
that seen for DNA or RNA in water. Unlike DNA where hybridized base 
pairs are radially opposite, in our PNA the hybridized bases are more 
likely to line up side by side in our attempts to model them. 
Furthermore, we have not been able to achieve an appropriate modeling 
of the balance between the hydrogen bond formation and the $\pi$ 
stacking between the bases in large part due to the hydrophobic and 
amphiphilic elements involved. More work and new ideas are needed 
here. However, we believe that the most fundamental properties of the 
biopolymer used---a PNA strand decorated with hydrophobic anchors 
that is able to hybridize with another PNA strand via H-bonds---is 
captured, at least in a qualitative manner. Against the background of 
this caveat, two findings are of particular interest: the simulations 
reveal that even our simple template representations are sufficient 
to introduce an impact on the stability of the hybridization complex. 
In other words, it is observed how a molecular fitness function 
emerges from very few assumptions about the underlying molecular 
implementation. Furthermore, this fitness function is not a simple 
superposition of the individual monomer properties, but rather 
depends on the \emph{sequence} of nucleotides in the genome. 

Equipartition of the components among the daughter cells after the 
division was achieved only when a few hydrophobic sensitizers are 
present in the protocell. Above a minimal number of sensitizers, 
equipartition becomes less probable as the number of sensitizers is 
further increased. This counter-intuitive finding is connected to the 
fact that sensitizers, like precursors, form a hydrophobic core in 
the interior of the micelle, thereby increasing the allowed size of 
stable aggregates, in addition to stabilizing them overall. Since the 
stability of the core itself increases with its size, once large 
enough, it becomes nearly impossible for the core and therefore the 
protocell as a whole to divide. Instead, the instability caused by 
the excess surfactants is addressed by rejecting excess individual 
surfactants one at a time. The 
results suggest that the volume of the sensitizer molecules most 
likely will affect the fission dynamics when a certain threshold is 
reached. 


Many open questions about systemic issues are still left unanswered 
by these initial investigations.  The main open issues include:
(i) What is the effect of heating the whole system in order to 
de-hybridize the gene templates? Obviously, the lipid aggregate has 
to be more heat tolerant than the gene duplex.  
(ii) What is the effect of defining the gene duplex as the 
photo-catalyst as in the originally proposed protocell design 
\citep{Ras:2003}? In our simulations, the sensitizer has been assumed 
to do the photo-fragmentation without any genetic 
catalysis.  Also, what is the effect of having the sensitizer as a 
separate molecule (as reported here) versus covalently linking it to 
the gene, e.g. as one of the lipophilic anchors?  
(iii) What is the effect on the overall protocell replication if both 
the gene precursors (oligomers) and the lipid precursors are supplied 
to the solution and have to diffuse to the protocell? In such a case,
will we see the coordinated gene and container growth based on
reaction kinetics predicted by \citet{Roc:2006}?
As gene replication is necessary before container division for two
viable daughters, can that be ensured in other ways than through a 
sequential resource supply? 
(iv) What new issues arise when the protocell goes through more 
than one generation of its life cycle, e.g. due to complementary 
resource sequence supplies? 

Subsequent work in this area must also relate the DPD simulation 
implementation in this publication and its dynamics with corresponding 
molecular dynamics simulations \citep{Wer:2006} and reaction kinetics 
studies \citep{Knu:2006} as well as experimental findings as they
arise.

\section{Conclusion}

The overall replication dynamics that constitute the life cycle of the
Los Alamos Bug was implemented using DPD simulations. In 
particular, we investigated the dynamics of container, metabolic 
complex, and genome subsystems, as well as the mutual interaction 
between these individual components. Component diffusion, 
self-assembly, precursor incorporation, metabolic turnover, template 
directed replication of the gene, and finally the protocellular 
division were studied in various simulations. The main systemic finds 
are: (a) Metabolic growth orchestration can be coordinated by a
switchable light source and/or by a continuous light source together
with regulation of the size and frequency of the oily 
precursor package injection, which was not anticipated. (b) As anticipated, 
there is a tradeoff between the lipophilic strength of the genetic 
backbone that makes it stick to the aggregate and its ability to 
easily hybridize with a complementary string. (c) As anticipated, for 
PNA with hydrophobic side chains, three dimensional structure 
formation that can potentially inhibit appropriate hybridization is 
more likely in water than at an oil-water or lipid-water interface,
although this is in part also dependent on the type pf hybridization
attraction. 
(d) Gene replication is easier at the surface of a micelle with a 
substantial oil core than for a micelle with a little or no oil core. 
The larger the oil core is, the easier the gene replication becomes
due to the aggregate stability and the ability to have a linear 
template. (e) As anticipated, the stability of two full 
complementary gene strings is much higher than a gene template and 
two complementary unligated gene pieces. (f) Rather surprisingly we 
observe that the template directed replication rate is dependent on 
the monomer component sequence and not only on the monomer component 
composition. (g)  Partition of lipids, 
sensitizers, and gene between daughter cells strongly depends on the 
size of the oil core. The smaller the oil core is, the more balanced 
the partition becomes, which was not anticipated. 

These systemic findings are now being considered in the experimental 
designs being pursued as part of the Protocell Assembly (PAs) and 
Programmable Artificial Cell Evolution (PACE) collaborations and 
their validity will eventually be addressed as the experiments are 
executed.

\begin{acknowledgments}
The authors would like to thank the members of the Barcelona Complex 
Systems Lab as well as members of the Los Alamos protocell team for 
useful discussions. This work is supported by the Programmable
Artificial Cell Evolution (PACE) project funded by the $6^{\mbox{th}}$ 
European Union Framework Program under contract FP600203 and the 
Los Alamos sponsored LDRD-DR Protocell Assembly (PAs) project. 
\end{acknowledgments}

\appendix
\section{Algorithm for chemical reactions}
\label{sec_algorithm}

Between every two DPD time steps, the following algorithm is applied
to perform chemical reactions:
For every reaction scheme, we successively check all possible pairs 
of reactants $A,B$, and compare their effective reaction rate $k$ to 
a number taken from a suitably normalized pseudo-random number 
generator. If the reaction rate is smaller than this value, we 
perform the reaction and go on to the next pair of possible reactants.
However, $A$ and $B$ will not be considered again in this step. The
exact algorithm---notated in the Python programming language---reads 
as follows:

\begin{footnotesize}
\begin{verbatim}
shuffle(reaction_list)
for reaction in reaction_list :
  for A in space.particles(reaction.educt_A) :

    if reaction.is_synthesis :
      # if reaction is a synthesis, possible
      # reaction partners are particles 
      # of type educt_B in the vicinity of A.
      partners = A.neighbors(
        reaction.educt_B,reaction.R
      )

    else :
      # otherwise, possible reaction partners
      # are particles of type educt_B bonded to A.
      partners = A.bonded(reaction.educt_B)
        
    for B in partners :
      # compute effective reaction rate
      k = reaction.k
      for C in A.neighbors(
        reaction.catalyst,reaction.r_cat
      ) :
        k += reaction.k_cat * 
          (1-(A.pos-C.pos).length()/reaction.r_cat)
                
        if random() < dt * k :
          # perform reaction
          react(A,B,reaction)
          # and leave loop over partners
          continue
\end{verbatim}
\end{footnotesize}

\bibliography{references}

\begin{thebibliography}{33}
\providecommand{\natexlab}[1]{#1}
\providecommand{\url}[1]{\texttt{#1}}
\expandafter\ifx\csname urlstyle\endcsname\relax
  \providecommand{\doi}[1]{doi: #1}\else
  \providecommand{\doi}{doi: \begingroup \urlstyle{rm}\Url}\fi

\bibitem[Alberts et~al.(2002)Alberts, Johnson, Lewis, Raff, Roberts, and
  Watson]{Alb:2002}
B.~Alberts, A.~Johnson, J.~Lewis, M.~Raff, K.~Roberts, and P.~Watson.
\newblock \emph{Molecular Biology of the Cell}.
\newblock Garland Science Publishing, 2002.

\bibitem[Aniansson et~al.(1976)Aniansson, Wall, Almgren, Hoffmann, Kielmann,
  Ulbricht, Zana, Lang, and Tondre]{Ani:1976}
E.~A.~G. Aniansson, S.~N. Wall, M.~Almgren, H.~Hoffmann, L.~Kielmann, W.~J.
  Ulbricht, R.~Zana, J.~Lang, and C.~Tondre.
\newblock Theory of the kinetics of micellar equilibria and quantitative
  interpretation of chemical relaxation studies of micellar solutions of ionic
  surfactants.
\newblock \emph{J. Phys. Chem.}, 80:\penalty0 905, 1976.

\bibitem[Bachmann et~al.(1992)Bachmann, Luisi, and Lang]{Bac:1992}
P.~Bachmann, P.~Luisi, and J.~Lang.
\newblock Autocatalytic self-replicating micelles as models for prebiotic
  structures.
\newblock \emph{Nature}, 357:\penalty0 57--59, 1992.

\bibitem[Bedau et~al.(2006)Bedau, Buchanan, Gozzala, Hanczyc, Maeke, McCaskill,
  Poli, and Packard]{Bed:2005}
M.~Bedau, A.~Buchanan, G.~Gozzala, M.~Hanczyc, T.~Maeke, J.~McCaskill, I.~Poli,
  and N.~Packard.
\newblock Evolutionary design of a {\sc ddpd} model of ligation.
\newblock In \emph{Proceedings of the 7th International Conference on
  Artificial Evolution EA'05}, pages 201--212. Springer, 2006.

\bibitem[Buchanan et~al.(2006)Buchanan, Gazzola, and Bedau]{Buc:2006}
A.~Buchanan, G.~Gazzola, and M.~A. Bedau.
\newblock \emph{Systems Self-Assembly: multidisciplinary snapshots}, chapter
  Evolutionary Design of a Model of Self-Assembling Chemical Structures.
\newblock Elsevier, 2006.

\bibitem[Coveney et~al.(1996)Coveney, Emerton, and Boghosian]{Cov:1996}
P.~V. Coveney, A.~N. Emerton, and B.~M. Boghosian.
\newblock Simulation of self-reproducing micelles using a lattice-gas
  automaton.
\newblock \emph{J. Amer. Chem. Soc.}, 118:\penalty0 10719--10724, 1996.

\bibitem[Espa{\~n}ol and Warren(1995)]{Esp:1995}
P.~Espa{\~n}ol and P.~Warren.
\newblock Statistical mechanics of dissipative particle dynamics.
\newblock \emph{Europhys. Lett.}, 30:\penalty0 191--196, 1995.

\bibitem[Evans and Wennerstr{\"o}m(1999)]{Eva:1999}
D.~Evans and H.~Wennerstr{\"o}m.
\newblock \emph{The Colloidal Domain - Where Physics, Chemistry, Biology, and
  Technology Meet}.
\newblock Wiley-VCH, New York, 1999.

\bibitem[Fellermann and Sol{\'e}(2006)]{Fel:2006}
H.~Fellermann and R.~Sol{\'e}.
\newblock Minimal model of self-replicating nanocells.
\newblock \emph{Phil. Trans. R. Soc. Lond. B}, 2006.
\newblock in press.

\bibitem[Ganti(2003)]{Gan:2003}
T.~Ganti.
\newblock \emph{The Principles of Life}.
\newblock Oxford University Press, 2003.

\bibitem[Groot(2000)]{Gro:2000}
R.~D. Groot.
\newblock Mesoscopic simulation of polymer-surfactant aggregation.
\newblock \emph{Langmuir}, 16:\penalty0 7493--7502, 2000.

\bibitem[Groot and Rabone(2001)]{Gro:2001}
R.~D. Groot and K.~L. Rabone.
\newblock Mesoscopic simulation of cell membrane damage, morphology change and
  rupture by nonionic sufactants.
\newblock \emph{Biophysical Journal}, 81:\penalty0 725--736, 2001.

\bibitem[Groot and Warren(1997)]{Gro:1997}
R.~D. Groot and P.~B. Warren.
\newblock Dissipative particle dynamics: Bridging the gap between atomistic and
  mesoscale simulation.
\newblock \emph{J. Chem. Phys.}, 107\penalty0 (11):\penalty0 4423--4435, 1997.

\bibitem[Hanczyc and Szostak(2004)]{Han:2004}
M.~Hanczyc and J.~Szostak.
\newblock Replicating vesicles as models of primitive cell growth and division.
\newblock \emph{Current Opinion in Chemical Biology}, 8:\penalty0 660--664,
  2004.

\bibitem[Hoogerbrugge and Koelman(1992)]{Hoo:1992}
P.~Hoogerbrugge and J.~Koelman.
\newblock Simulating microscopic hydrodynamic phenomena with dissipative
  particle dynamics.
\newblock \emph{Europhys. Lett.}, 19:\penalty0 155--160, 1992.

\bibitem[Knutson et~al.(2006)Knutson, Benk{\"o}, Rocheleau, Mouffouk, Maselko,
  Shreve, Chen, and Rasmussen]{Knu:2006}
C.~Knutson, G.~Benk{\"o}, T.~Rocheleau, F.~Mouffouk, J.~Maselko, A.~Shreve,
  L.~Chen, and S.~Rasmussen.
\newblock Metabolic photo-fragmentation kinetics for a minimal protocell.
\newblock \emph{Artifical Life}, 2006.

\bibitem[Luisi et~al.(1994)Luisi, Walde, and Oberholzer]{Lui:1994}
P.~Luisi, P.~Walde, and T.~Oberholzer.
\newblock Enzymatic {RNA} synthesis in selfreproducing vesicles: An approach to
  the construction of a minimal synthetic cell.
\newblock \emph{Ber. Bunsenges. Phys. Chem.}, 98:\penalty0 1160--1165, 1994.

\bibitem[Marsh(1998)]{Mar:1998}
C.~Marsh.
\newblock \emph{Theoretical Aspects of Dissipative Particle Dynamics}.
\newblock PhD thesis, Lincoln College, University of Oxford, 1998.

\bibitem[Mayer and Rasmussen(1998)]{May:1998}
B.~Mayer and S.~Rasmussen.
\newblock Self-reproduction of dynamical hierarchies in chemical systems.
\newblock In \emph{Artificial Life VI Proceedings}, pages 123--129. MIT Press,
  1998.

\bibitem[Mayer and Rasmussen(2000)]{May:2000}
B.~Mayer and S.~Rasmussen.
\newblock Dynamics and simulation of micellular self-reproduction.
\newblock \emph{Int. J. Mod. Phys. C}, 11:\penalty0 809--826, 2000.

\bibitem[McCaskill et~al.(2006)McCaskill, Packard, Rasmussen, and
  Bedau]{Cas:2006}
J.~McCaskill, N.~Packard, S.~Rasmussen, and M.~Bedau.
\newblock Evolutionary self-organization in complex fluids.
\newblock \emph{Phil. Trans. R. Soc. Lond. B}, 2006.
\newblock in press.

\bibitem[Nielson(2007)]{Nie:2007}
P.~Nielson.
\newblock \emph{Protocells: Brdiging nonliving and living mater}, chapter
  Peptide nucleic acid ({PNA}) as prebiotic and abiotic genetic material.
\newblock MIT Press, 2007.
\newblock in press.

\bibitem[Ono(2001)]{Ono:2001}
N.~Ono.
\newblock \emph{Artificial Chemistry: Computational Studies on the Emergence of
  Self-Reproducing Units}.
\newblock PhD thesis, Institute of Physics, University of Tokyo, 2001.

\bibitem[{PAs \& PACE}(2004-2008)]{PAsPACE}
{PAs \& PACE}, 2004-2008.
\newblock An integrated multiscale computational and experimental approach is
  applied in the Los Alamos National Laboratory sponsored Protocell Assembly
  (PAs) project and the European Commission sponsored Programmable Artificial
  Cell Evolution (PACE) project.

\bibitem[Pohorille and Deamer(2002)]{Poh:2002}
A.~Pohorille and D.~Deamer.
\newblock Artificial cells: prospects for biotechnology.
\newblock \emph{Trends in Biotechnology}, 20:\penalty0 123--128, 2002.

\bibitem[Rasmussen et~al.(2003)Rasmussen, Chen, Nilsson, and Abe]{Ras:2003}
S.~Rasmussen, L.~Chen, M.~Nilsson, and S.~Abe.
\newblock Bridging nonliving and living matter.
\newblock \emph{Artificial Life}, 9:\penalty0 269--316, 2003.

\bibitem[Rocheleau et~al.(2006)Rocheleau, Rasmussen, Nielson, Jacobi, and
  Ziock]{Roc:2006}
T.~Rocheleau, S.~Rasmussen, P.~E. Nielson, M.~N. Jacobi, and H.~Ziock.
\newblock Emergence of protocellular growth laws.
\newblock \emph{Phil. Trans. R. Soc. B}, 2006.
\newblock in press.

\bibitem[Trofimov(2003)]{Tro:2003}
S.~Trofimov.
\newblock \emph{Thermodynamic consistency in dissipative particle dynamics}.
\newblock PhD thesis, Technische Universiteit Eindhoven, 2003.

\bibitem[Venturoli and Smit(1999)]{Ven:1999}
M.~Venturoli and B.~Smit.
\newblock Simulating self-assembly of model membranes.
\newblock \emph{Phys. Chem. Comm.}, 10, 1999.

\bibitem[Weronski et~al.(2006)Weronski, Jiang, and Rasmussen]{Wer:2006}
P.~Weronski, Y.~Jiang, and S.~Rasmussen.
\newblock Molecular dynamics ({MD}) study of small {PNA} molecule in
  lipid-water system.
\newblock \emph{J. Biophys.}, 2006.
\newblock in press.

\bibitem[Yamamoto and Hyodo(2003)]{Yam:2003}
S.~Yamamoto and S.~Hyodo.
\newblock Budding and fission dynamics of two-component vesicles.
\newblock \emph{J. Chem. Phys.}, 118\penalty0 (17):\penalty0 7937--7943, 2003.

\bibitem[Yamamoto et~al.(2002)Yamamoto, Maruyama, and Hyodo]{Yam:2002}
S.~Yamamoto, Y.~Maruyama, and S.~Hyodo.
\newblock Dissipative particle dynamics study of spontaneous vesicle formation.
\newblock \emph{J. Chem. Phys.}, 116\penalty0 (13):\penalty0 5842--5849, 2002.

\bibitem[Yuan et~al.(2002)Yuan, Cai, and Xu]{Yua:2002}
S.~Yuan, Z.~Cai, and G.~Xu.
\newblock Dynamic simulation of aggregation in surfactant solution.
\newblock \emph{Acta Chimica Sinica}, 60\penalty0 (2):\penalty0 241--245, 2002.

\end{thebibliography}

\end{document}